\documentclass[11pt]{article}
\usepackage[utf8]{inputenc}
\usepackage[a4paper, top=4cm, bottom=4cm, left=3cm, right=3cm]{geometry}
\usepackage{amsmath, amssymb, amsthm}
\usepackage{color}
\usepackage{todonotes}
\usepackage{url}
\usepackage[colorlinks, allcolors=blue]{hyperref}
\usepackage{enumitem}
\usepackage{graphicx}
\usepackage{subcaption}

\newtheorem{theorem}{Theorem}[section]
\newtheorem{corollary}[theorem]{Corollary}
\newtheorem{lemma}[theorem]{Lemma}
\newtheorem{claim}[theorem]{Claim}

\theoremstyle{definition}
\newtheorem{definition}[theorem]{Definition}
\newtheorem{example}[theorem]{Example}

\theoremstyle{remark}
\newtheorem{remark}[theorem]{Remark}

\newenvironment{claimproof}{\begin{proof}\renewcommand{\qedsymbol}{\claimqed}}{\end{proof}\renewcommand{\qedsymbol}{\plainqed}}
\let\plainqed\qedsymbol

\DeclareMathOperator{\sgon}{sgon}
\DeclareMathOperator{\tw}{tw}
\DeclareMathOperator{\indeg}{indeg}
\DeclareMathOperator{\outdeg}{outdeg}
\DeclareMathOperator{\width}{b}

\newcommand{\Z}{\mathbb{Z}}

\newcommand{\FPT}{\textup{FPT}} 
\newcommand{\W}{\textup{W$[1]$}} 
\newcommand{\XNLP}{\textup{XNLP}}

\title{Problems hard for treewidth\\ but easy for stable gonality}
\author{Hans L.\ Bodlaender\thanks{Department of Information and Computing Sciences, Utrecht University, Princetonplein 5, 3584CC Utrecht, the Netherlands, h.l.bodlaender@uu.n } \and Gunther Cornelissen\thanks{Department of Mathematics,
Utrecht University,
P.O.~Box 80010, 3508 TA Utrecht,The Netherlands, g.cornelissen@uu.nl} \and
Marieke van der Wegen\thanks{Department of Mathematics,
Utrecht University,
P.O.~Box 80010, 3508 TA Utrecht,The Netherlands, m.vanderwegen@uu.nl}}
\date{\today}

\begin{document}

\maketitle

\begin{abstract}\noindent We show that some natural problems that are \XNLP-hard (which implies $\mathrm{W}[t]$-hardness for all $t$) when parameterized by pathwidth or treewidth, become FPT when parameterized by stable gonality, a novel graph parameter based on optimal maps from graphs to trees. 
The problems we consider are classical flow and orientation problems, such as  \textsc{Undirected Flow with Lower Bounds} (which is strongly {NP}-complete, as shown by Itai), \textsc{Minimum Maximum Outdegree} (for which $\mathrm{W}[1]$-hardness for treewidth was proven by Szeider), and capacitated optimization problems such as \textsc{Capacitated (Red-Blue) Dominating Set} (for which $W[1]$-hardness was proven by Dom, Lokshtanov,  Saurabh and Villanger).  
Our hardness proofs (that beat existing results) use reduction to a recent \XNLP-complete problem (\textsc{Accepting Non-deterministic Checking Counter Machine}). The new, ``easy'', parameterized algorithms use a novel notion of weighted tree partition with an associated parameter that we call ``treebreadth'', inspired by Seese's notion of tree-partite graphs, as well as techniques from dynamical programming and integer linear programming. 
\end{abstract}

\section{Introduction}

\paragraph*{The parameterization paradigm} Problems on finite (multi-)graphs that are NP-hard may become polynomial by restricting a specific graph parameter $k$. If furthermore the degree of the resulting polynomial bound is constant, we say that the problem becomes \emph{fixed parameter tractable} (FPT) for the parameter $k$ \cite[1.1]{PA}. More precisely, there should exist an algorithm that solves the given problem in time bounded by a computable function of the parameter $k$ times a power of the input size of the problem. Despite the fact that computing the parameter itself can often be shown to be NP-hard or NP-complete, the FPT-paradigm, originating in the work of Downey and Fellows \cite{DF}, has shown to be very fruitful both in theory and practice. 

One  successful approach is to consider graph parameters that measure how far a given graph is from being acyclic; more specifically, how the graph may be decomposed into ``small'' pieces, such that the interrelation of the pieces is described by a tree-like structure. A prime example of such a parameter is the \emph{treewidth} $\tw(G)$ of a graph $G$ (\cite[Ch.\ 7]{PA}; reviewed briefly below). Some reasons for the success are that graphs of bounded treewidth are amenable to algorithms based on dynamic programming \cite[7.3]{PA}, and Courcelle's theorem \cite{Courcelle} \cite[7.4]{PA}, that states that graph problems described in second order monadic logic are FPT, linear in the number of vertices, provided a tree decomposition realising the treewidth is given. Examples of problems that are FPT for treewidth are \textsc{Dominating Set} and \textsc{Vertex Cover} \cite[7.3]{PA}. 

Other parameters have been considered; for example, in \cite{FGK}, it is shown that colouring problems such as \textsc{Equitable Colouring} (existence of a vertex colouring with at most $k$ colours, so that the sizes of any two colour classes differ by at most one) is hard for treewidth (technically, $\mathrm{W}[1]$-hard), but FPT for vertex cover number. In \cite{beyond} and \cite[7.9]{PA} one finds more problems and parameters that shift their complexity. Still, some famous graph orientation and graph flow problems, as well as capacitated version of classical problems such as \textsc{Dominating Set} have so far not succumbed to the \FPT\ paradigm for any reasonable parameter, despite being of practical importance in logistics and resource allocation. Below, we will consider \textsc{Undirected Flow with Lower Bounds}, Problem [ND37] in \cite{GareyJ79}, known to be strongly NP-complete since the 1970's. We will show that, when parameterized by treewidth, the problem is hard (in fact, \XNLP-complete with pathwidth as parameter, cf.\ infra). The new paradigm of this paper is that such problems become \FPT\ for a novel, natural graph parameter based on \emph{mapping} the graph to a tree, rather than decomposing the graph. 

\paragraph*{A novel parameter: stable gonality} This novel multigraph parameter, based on ``tree-likeness'', is the so-called \emph{stable gonality} $\sgon(G)$ of a multigraph $G$, introduced in \cite[\S 3]{CKK}. As we will briefly describe in in Remark \ref{rem:RS}, the parameter originates in algebraic geometry -- more precisely, the theory of Riemann surfaces -- where a similar construction has been used since the 19th century; the analogy between Riemann surfaces and graphs stretches much further, as seen, for example, in \cite{Baker} and \cite{BakerNorine}. The basic idea is to replace the use of tree \emph{decompositions} of a graph $G$ by graph \emph{morphisms} from $G$ to trees, and to replace the ``width'' of the decomposition by the ``degree'' of the morphism, where lower degree maps correspond to less complex graphs. For example, graphs of stable gonality $1$ are trees \cite[Example 2.13]{hyperelliptic}, those of stable gonality $2$ are so-called hyperelliptic graphs, i.e., graphs that admit, after refinement, a graph automorphism of order two such that the quotient graph is a tree; this is decidable in quasilinear time \cite[Thm.\ 6.1]{hyperelliptic}. There are two technicalities that we will explain in the body of the text, but ignore for now: (a) in order to be able to define a notion of ``degree'' of a graph morphism, one needs to assume that the map is \emph{harmonic} for a certain choice of edge weights; (b) it turns out that the most useful parameter occurs by further allowing refinements of the graph (this explain the terminology ``stable'', a loanword from algebraic geometry). A detailed definition with intuition and examples is given in in Section \ref{sec:decompositiontree}.

It has been shown that $\tw(G) \leq \sgon(G)$  \cite[\S 6]{twsgon} and $\sgon(G) \leq (c(G)+3)/2$, where $c(G)$ is the cyclomatic number of $G$ \cite[Thm.\ 5.7]{CKK}, that $\sgon(G)$ is computable, and NP-complete \cite{gon-hard, sgon-computable}. One attractive point of stable gonality as parameter for weighted problems stems from the fact that it is sensitive to multigraph properties, whereas the treewidth of a multigraph equals that of the underlying simple graph. Just like for treewidth \cite{SCS}, lower bounds are known, depending on the Laplace spectrum and maximal degree of the graph, leading, for example, to lower bounds for the stable gonality of expanders linear in the number of vertices \cite[Cor.\ 6.10]{CKK}. 

\paragraph*{Three sample problems} As stated above, the goal of this paper is to show that certain problems that are hard in bounded treewidth become easy (in fact, \FPT) for stable gonality. In this introduction, we discuss three examples (one about orientation, one about flow, and one on capacitated domination), but in the body of the paper we consider many more related problems. We always assume that integers are given in unary.

\subparagraph*{\underline{An orientation problem}} A typical orientation problem is the following. 
 
\begin{verse}
{\sc Minimum Maximum Outdegree} (cf.\ Szeider \cite{Szeider11}) \\
{\bf Given:} Undirected weighted graph $G=(V,E,w)$ with a weight function $w \colon E \rightarrow {\Z}_{>0}$; integer $r$\\
{\bf Question:} Is there an orientation of $G$ such that for each $v\in V$, the total weight of all edges directed out of $v$ is at most $r$?
\end{verse}

This and related problems naturally concern weighted graphs, and we define their stable gonality in terms of an associated multigraph: given an undirected weighted graph $G=(V,E,w)$, we have an associated (unweighted) multigraph $\tilde G$, with the same vertex set, but where each simple edge $e=uv$ in $G$ is replaced by $w(e)$ parallel edges between the vertices $u$ and $v$. We call the stable gonality of the associated multigraph $\tilde G$ the stable gonality of the weighted graph $G$, and denote it by $\sgon(G):=\sgon(\tilde G)$.  

\subparagraph*{\underline{A flow problem}}  To describe a typical flow problem, we first recall some notions from the theory of network flow. A \emph{flow network} is a \emph{directed} graph $D=(N,A)$ with for each
arc $e \in A$ a capacity $c(e) \in \Z_{>0}$, and two nodes $s$ (source) and $t$ (target) in $N$. Given a function $f\colon A \rightarrow \Z_{\geq 0}$ and a node $v$, we call $\sum_{wv\in A} f(wv)$ the {\em flow to} $v$ and $\sum_{vw\in A} f(vw)$ the {\em flow out of} $v$. We say $f$ is an
{\em $s$-$t$-flow} if for each arc $a\in A$, the flow over the arc is nonnegative and at most its capacity (i.e., $0\leq f(a) \leq c(a)$), and for each node $v\in N\setminus \{s,t\}$, the flow conservation law holds: the flow to $v$ equals the flow out of $v$. 
The {\em value} $\mathrm{val}(f)$ of a flow is the flow out of $s$ minus the flow to $s$. A flow $f$ is a {\em circulation} if it has value $\mathrm{val}(f)=0$, which is equivalent to 
flow conservation also holding at $s$ and $t$.

\begin{verse}
 {\sc Undirected Flow with Lower Bounds} \\
 {\bf Given:} Undirected graph $G=(V,E)$, for each edge $e\in E$ a positive integer capacity $c(e) \in \Z_{>0}$ and a non-negative integer lower bound $\ell(e) \in \Z_{\geq 0}$, vertices $s$ (source) and $t$ (target), a non-negative integer $R \in \Z_{>0}$ (value) \\
 {\bf Question:} Is there an orientation of $G$ such that the resulting directed graph $D$ allows an $s$-$t$-flow $f$ that meets capacities and lower bounds (i.e., $\ell(a) \leq f(a) \leq c(a)$ for all arcs in $D$), with value $R$? 
\end{verse}

This is Problem [ND37] in the classical reference Garey and Johnson \cite{GareyJ79} (in that reference it is required that $\mathrm{val}(f) \geq R$ rather than $\mathrm{val}(f) = R$, but the problems are seen to be of the same complexity by adding a new target vertex $t'$ with edge $tt'$.
For the transformation in one direction, set $\ell(tt')=c(tt')=R$; in the other direction, set $\ell(tt')=R$ and choose $c(tt')$ sufficiently large, e.g., equal to the sum of the weights of all edges incident to $t$ in $G$, cf.\ \cite{Itai}.)

The corresponding problem on a \emph{directed} graph is solvable in polynomial time. 

\subparagraph*{\underline{A capacitated problem}}  Capacitated version of classical graph problems concern imposing a limitation on the available ``resources'', placing them closer to real-world situations. Our final type of problem is a capacitated version of \textsc{Dominating Set} (to find out whether there is a set of vertices of size $\leq k$ that is connected to all vertices in the graph), that can be viewed as an abstract form of facility location questions.

\begin{verse}
 {\sc Capacitated Dominating Set} \\
 {\bf Given:} Undirected graph $G=(V,E)$, for each vertex $v\in V$ a positive integer capacity $c(v) \in \Z_{>0}$, integer $k$ \\
 {\bf Question:} Is there a set $D \subset V$ of size $|D| \leq k$ and a function $f \colon V \setminus D \rightarrow D$ such that $vf(v) \in E$  for all $v\in V\setminus D$ and $|f^{-1}(v)| \leq c(v)$ for all $v \in D$?   
\end{verse}

\paragraph*{Complexity classes} To make the (parameterized) hardness of these problems more precise, we use the complexity class \XNLP\ parameterized by a graph parameter $k$,  first considered by Elberfeld, Stockhusen and Tantau in \cite{ElberfeldST15}; this is the class of problems that can be solved non-deterministically in time $O(f(k)n^c)$ ($c \geq 0$) and space $O(f(k) \log(n))$ where $n$ is the input size and $f$ is a computable function. More familiar is the W-hierarchy of Downey and Fellows up to parameterised reduction \cite[13.3]{PA}, where $\mathrm{W}[0]$ is \FPT, and \W\ is a class for which \textsc{Independent Set} is complete (parameterized by the size of the set). We presently note that \XNLP-hardness implies $\mathrm{W}[t]$-hardness for all $t$ (for a given parameter) \cite[Lemma 2.2]{BodlaenderGNS21}. 

 \paragraph*{Main result: hard problems for treewidth become easy for stable gonality} 
 
\begin{theorem} \textsc{Minimum Maximum Outdegree}, \textsc{Undirected Flow with Lower Bounds} and \textsc{Capacitated Dominating Set} are 
\XNLP-complete for pathwidth, and \XNLP-hard for treewidth \textup{(}given a path or tree decomposition realising the path- or treewidth\textup{)}, but are \FPT\ for stable gonality \textup{(}given a refinement and graph morphism from the associated multigraph to a tree realising the stable gonality\textup{)}.
\end{theorem}  

Unparameterized {\sc Undirected Flow with Lower Bounds} has been known since 1977  to be NP-complete in the strong sense, cf.\ Itai \cite[Theorem 4.1]{Itai}, \cite[p.\ 216]{GareyJ79}). Uncapacitated \textsc{Dominating set} is $W[2]$-complete for the size of the dominating set  \cite[Theorem 13.28]{PA}, and \FPT\ for treewidth \cite[Theorem 7.7]{PA}. 
The \W-hardness of \textsc{Minimum Maximum Outdegree} for treewidth was proven by Szeider \cite{Szeider11} and of \textsc{Capacitated Dominating Set} by Dom, Lokshtanov,  Saurabh and Villanger \cite{Dom}.

\XNLP-completeness of \textsc{Capacitated Dominating Set} for pathwidth  in the main theorem is not due to us, but was proven very recently by  Bodlaender, Groenland and Jacob building upon our results,  see \cite[Theorem 8]{BodlaenderGJ22}. 

As far as we know, our proof is the first parameterized-hardness result for \textsc{Undirected Flow with Lower Bounds}. 
We prove the $\XNLP$-hardness for pathwidth (and hence, for treewidth) by reduction from \textsc{Accepting Non-deterministic Checking Counter Machine} from \cite{BodlaenderGNS21}. see Section \ref{sec:hardness}.  

For proving \FPT\ under stable gonality, we revive an older idea of Seese on tree-partite graphs and their widths \cite{Seese};  in contrast to the tree decompositions used in defining treewidth, we partition the original graph vertices into \emph{disjoint} sets (`bags') labelled by vertices of a tree, such that adjacent vertices are in the same bag or in bags labelled by adjacent vertices in the tree. Seese introduced \emph{tree partition width} to be the maximal size of a bag in such any partition. We consider weighted graphs and define a new parameter, \emph{breadth}, given as the maximum of the bag size and the sum of the weights of edges between adjacent bags; cf.\ Subsection \ref{sec:decompositiontree} below, in particular, Figure \ref{breadthfig} for a schematic illustration. Taking the minimal breadth over all tree partitions gives a new graph parameter, that we call \emph{treebreadth.} This allows us to divide the proof in two parts: (a) show that, given a graph morphism from the associated multigraph to a tree, one can compute in polynomial time a tree partition of the weighted graph of breadth upper bounded by the stable gonality of the associated multigraph --- see Theorem \ref{fhm-to-dt}; (b) provide an \FPT-algorithm,  given a tree partition of bounded breadth.  The very general intuition, to be made precise in the detailed proofs, is that the tree that we map the graph to (or that labels the bags in the tree partition) ``structures'' the algorithm by consecutively running bounded algorithms over the pre-images of  individual vertices and edges in the tree, using dynamical programming and integer linear programming to control extension of partial solutions to the entire graph. We mix this with classical tools for equivalence of various flow problems, and Edmonds' 
algorithm for matching. By reductions, the two algorithms we specify are described in the following theorem for the indicated parameters. 

\begin{theorem}  \textsc{Minimum Maximum Outdegree} is \FPT\ for treebreadth \textup{(}given a partition tree realising the treebreadth\textup{)}, and \textsc{Capacitated Dominating Set} is \FPT\ for tree partition width \textup{(}given a tree partition with bounded width\textup{)}. 
\label{theorem:fpt}
\end{theorem}  

We underline that Seese's original tree partition width suffices for the capacitated problem, by an argument described at the start of Subsection \ref{section:Cds-tech}.

The algorithm for \textsc{Minimum Maximum Outdegree} is given in Section \ref{section:algorithm} (in fact, for the closely related problem {\sc Outdegree Restricted Orientation}, in which outdegree is required to belong to a given interval instead of not exceeding a given value), and the algorithm for \textsc{Capacitated Dominating Set} is given in Section \ref{section:dominatingset}. 

In a related direction, we may weaken the complexity class but increase the strength of the parameter. Here, we prove the following, by reduction from {\sc Bin Packing} \cite{JansenKMS13}.

\begin{theorem} 
 \textsc{Minimum Maximum Outdegree} and \textsc{Undirected Flow with Lower Bounds} are \W-hard for vertex cover number. 
\end{theorem}  

\paragraph*{Further problems} In Section \ref{sec:problems}, we define some more related circulation problems (sometimes used as intermediaries in our arguments) and show that they have the same properties; these concern finding orientations on weighted graphs that make the weights define a circulation; or with the outdegree belonging to a given set, having a given value, or not exceeding a given value; and finding an $s$-$t$-flow on a given directed graph for which all non-zero values on arcs match the capacity exactly; and, finally, a coloured version of \textsc{Capacitated Dominating Set}.

\section{Preliminaries} 

\subsection{Conventions and notations}
We will consider \emph{multigraphs}, where we allow for parallel edges and self-loops. Said otherwise, a multigraph $G=(V,E)$ consists of a finite set $V$ of vertices, as well as a finite multiset $E$ of unoriented (unweighted) edges, i.e., a set of pairs of (possibly equal) vertices, with finite multiplicity on each such pair. We denote such an edge between vertices $u,v \in V$ as $uv$. For $v \in V$, $E_v$ denotes the edges incident with $v$, and for two subsets $X,Y \subset V$, $E(X,Y)$ is the collection of edges from any vertex in $X$ to any vertex in $Y$. 

We also consider \emph{weighted simple graphs}, where edges have positive integer weights. We will make repeated use of the correspondence between integer weighted simple graphs and multigraphs given by replacing every edge with weight $k$ by $k$ parallel edges. 

All graphs we consider are connected. For convenience, we use the terminology ``vertex'' and ``edge'' for undirected graphs, and ``arc'' and ``node'' for either directed graphs, or for trees that occur in graph morphisms or tree partitions.

We write $\Z$ for all integers, with unique subscripts indicating ranges (so $\Z_{>0}$ is the positive integers and $\Z_{\geq 0}$ the non-negative integers). We use interval notation for sets of integers, e.g., $[2,5] = \{2,3,4,5\}$.

\subsection{Stable gonality and treebreadth}
\label{sec:decompositiontree}

\paragraph*{Stable gonality}
The notion of stable gonality of a multigraph is a measure of tree-likeness of a multigraph defined using the minimal ``degree'' of a \emph{map} to a tree, rather than the more conventional ``decomposition'' in terms of trees. As usual, a \emph{graph homomorphism} between two loopless multigraphs $G$ and $H$, denoted $\phi\colon G \to H$, is a map of vertices that respects the incidences given by the edges, i.e., it consists of two (not necessarily surjective) maps  $\phi\colon V(G) \rightarrow V(H)$ and $\phi\colon E(G) \rightarrow E(H)$ such that $\phi(uv)  = \phi(u)\phi(v) \in E(H)$ for all $uv\in E(G)$. 
One would like to define the ``degree'' of such a graph homomorphism as the number of pre-images of any vertex or edge, but in general, this obviously depends on the chosen vertex or edge. However, by introducing certain weights on the edges via an additional index function, we get a large collection of ``indexed'' maps for which the degree can be defined as the sum of the indices of the pre-image of a given edge, as long as the indices satisfy a certain condition of ``harmonicity'' above every vertex in the target. We make this precise in the following definition.  

\begin{definition}
A \emph{finite morphism} $\phi$ between two loopless multigraphs $G$ and $H$ consists of a graph homomorphism $\phi\colon G \to H$ (denoted by the same letter), and an index function $r\colon E(G) \to \Z_{>0}$ (hidden from notation). The \emph{index of $v \in V(G)$ in the direction of $e \in E(H)$}, where $e$ is incident to $\phi(v)$, is defined by 
\begin{align*}
    m_{e}(v) := \sum_{\substack{e' \text{ incident to } v,\\ \phi(e') = e}} r(e').
\end{align*}
We call $\phi$ \emph{harmonic} if this index is independent of the direction $e \in E(H)$ for any given vertex $v \in V(G)$. We call this simply the \emph{index of $v$}, and denote it by $m(v)$. 
The \emph{degree} of a finite harmonic morphism $\phi$ is \begin{align*}
    \deg(\phi) = \sum_{\substack{e'\in E(G),\\ \phi(e') = e}} r(e') = \sum_{\substack{v'\in V(G),\\ \phi(v') = v}} m(v').
\end{align*}
where $e \in E(H)$ is any edge and $v \in V(H)$ is any vertex. Since $\phi$ is harmonic, this number does not depend on the choice of $e$ or $v$, and both expressions are indeed equal. 
\end{definition} 

The second ingredient in the definition of stable gonality is that of a refinement. 

\begin{definition}
Let $G$ be a multigraph. A \emph{refinement} of $G$ is a graph obtained using the following two operations iteratively finitely often:
\begin{itemize}
    \item add a leaf (i.e., a vertex of degree one), 
    \item subdivide an edge. 
\end{itemize}
\end{definition}

\begin{definition}
Let $G$ be a multigraph. The \emph{stable gonality} of $G$ is
\begin{align*}
    \sgon(G) = \min\{\deg(\phi) \mid \;&\phi\colon H\to T, \text{a finite harmonic morphism, where}\\
    & H \text{ is a loopless refinement of } G \text{ and} \\
    & T \text{ is a tree} \}.
\end{align*}
\end{definition}

\begin{figure}[t]
	\centering
	\begin{subfigure}{.45\textwidth}
		\centering
		\includegraphics{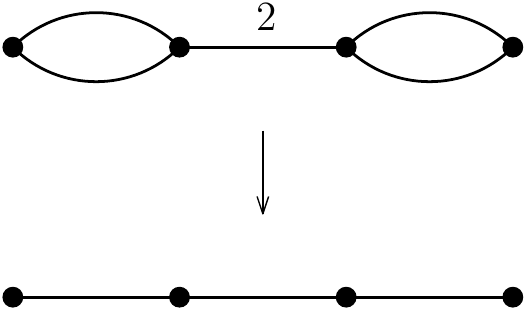}
		\caption{}
	\end{subfigure}
	\begin{subfigure}{.45\textwidth}
		\centering
		\includegraphics{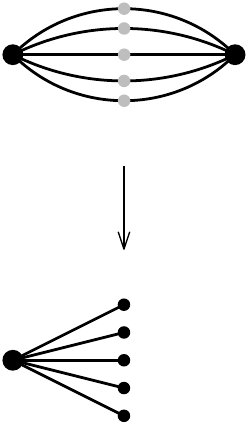}
		\caption{}
	\end{subfigure}
	\caption{Two examples of a finite harmonic morphisms of degree $2$. The edges without label have index $1$. The small grey vertices represent refinements of the graph. }
	\label{fig:ex-morphism}
\end{figure}

\begin{example} 
Two examples are found in Figure \ref{fig:ex-morphism}. The left-hand side illustrates the need for an index function (the middle edge needs label $2$), and the right hand side shows the effect of subdivision (without subdivision, the minimal degree to a tree is the total number of edges). Since the top multigraphs are not trees, and trees are precisely the multigraphs having stable gonality $1$ (\cite[Example 2.13]{hyperelliptic}), we conclude that they both have stable gonality two.
\end{example} 

\begin{remark} \label{rem:RS} 
The notion is similar to that of the gonality of a compact Riemann surface (or smooth projective algebraic curve), which is defined as the minimal degree of a non-constant holomorphic map to the Riemann sphere (or the projective line; the unique Riemann surface of first Betti number $0$, so the analogue of a tree). The need for refinements in the definition of stable gonality of graphs reflects the fact that there are infinitely many trees, whereas there is only one compact Riemann surface of first Betti number $0$. 
\end{remark}

\paragraph*{Tree partitions and their breadth} 
The existence of a harmonic morphism to a tree imposes a special structure on the graph that we can exploit in designing algorithms. To capture this structure, we define the ``breadth'' of \emph{tree partitions} of weighted graphs. The notion resembles that of ``tree-partite graphs'', introduced by Seese~\cite{Seese}. The idea is to \emph{partition} a (weighted) graph according to an index set given by the vertices of a tree, and use the incidence relations on the tree to define an associated measure for the graph. 

\begin{definition} A \emph{tree partition} $\mathcal T$ of a weighted graph $G=(V,E,w)$ is a pair
\begin{align*}\mathcal T=(\{X_i \mid i\in I\},\  T=(I,F))
\end{align*}
where $X_i$ are subsets of the vertex set $V$ and $T=(I,F)$ is a tree, such that 
$\{X_i \mid i\in I\}$ forms a partition of $V$ (i.e., for each $v\in V$, there is exactly one $i\in I$ with $v\in X_i$); and
adjacent vertices are in the same set $X_i$ or in sets corresponding to adjacent nodes (i.e., for each $uv \in E$, there exists an $i \in I$ such that $\{u,v\} \subseteq X_i$ or there exists $ij \in F$ with $\{u,v\} \subseteq X_i \cup X_j$).

The \emph{breadth}  
of a partition tree $\mathcal T$ of $G$ is defined as
\begin{align*}
    \width(\mathcal T):= \max \{ |X_i|, |E(X_j,X_k)| \mid i \in I, jk \in F \},
\end{align*}
with $$|E(X_j,X_k)| = \sum\limits_{e \in E(X_j,X_k)} w(e)$$ the weighted number of edges connecting vertices in $X_j$ to vertices in $X_k$.
\end{definition} 

\begin{figure}[t]
	\begin{center}	\includegraphics{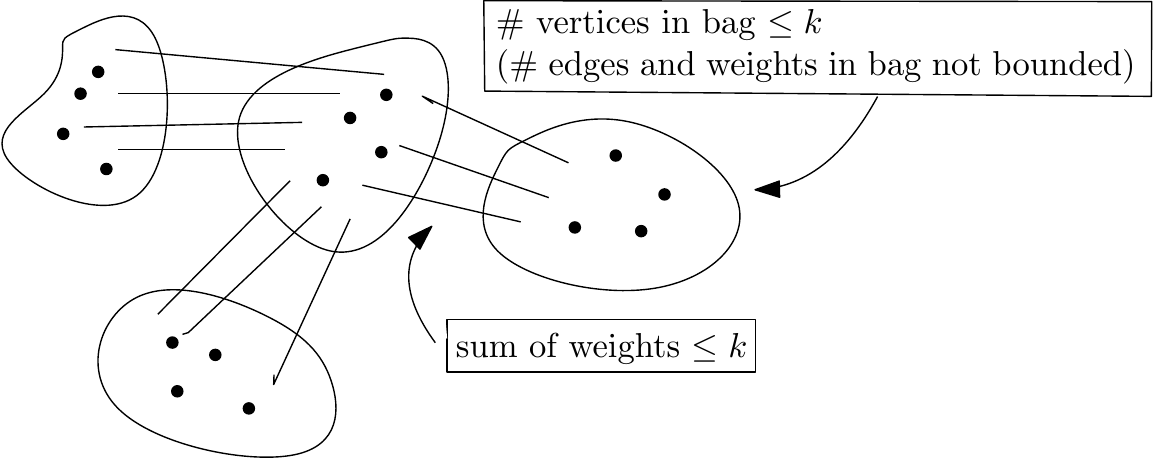} \end{center} 
	\caption{Schematic representation of a tree partition of a graph of breadth $\leq k$ }
	\label{breadthfig}
\end{figure}

We refer to Figure \ref{breadthfig} for a schematic view of a tree partition with weights and bounded breadth. 

If we have a tree partition of a weighted graph $G$ using a tree $T$, for convenience we will call the vertices of $T$ \emph{nodes} and the edges of $T$ \emph{arcs}. We call the sets $X_i$ \emph{bags}. Observe that in a tree partition of breadth $k$, if there are more than $k$ parallel edges between two vertices $u$ and $v$, then $u$ and $v$ will be in the same bag. 

\begin{definition} The \emph{treebreadth} $\mathrm{tb}(G)$ of a weighted graph $G$ is the minimum breadth of a tree partition of $G$. The \emph{stable treebreadth} $\mathrm{stb}(G)$ of a graph $G$ is the minimum treebreadth of any refinement of $G$. 
\end{definition} 

\begin{remark} 

By contrast to tree partitions, a \emph{tree decomposition} of a simple graph $G$ is a pair $\mathcal T=(\{X_i~|~i\in I\}, T=(I,F))$ such that $\bigcup X_i = V$ \emph{but the $X_i$ are not necessarily disjoint}, where for every $uv \in E$, $\{u,v\} \subseteq X_i$ for some $i \in I$, and for any $v \in V$ the set of vertices $i \in I$ such that $v \in X_i$ induces a connected subtree of $T$. The width of $\mathcal T$ is $\max \{ |X_i|-1 \mid i \in I\}$, and \emph{treewidth} is the minimal width of all tree decompositions. For \emph{pathwidth}, one furthermore insists that $T$ is a path graph. 

In Seese's work \cite{Seese}, the structure/weights of edges between bags does not contribute to the total width; Seese's tree-partition-width $\mathrm{tpw}(G)$ of a simple graph $G$, defined as the minimum over all tree partitions of $G$ of the maximum bag size in the tree partition, 
is thus a lower bound for the treebreadth $\mathrm{tb}(G)$ (in particular, for $\sgon$, cf.\ infra). For any $G$, $\mathrm{tpw}(G)$ is lower bounded in terms of $\tw(G)$, but also upper bounded in terms of $\tw(G)$ and the maximal degree in $G$, cf.\ \cite{Wood}. 
\end{remark} 

\paragraph*{From morphisms to tree partitions} 
\label{subsec:morphism-to-decomposition}

There is a direct relation between the existence of a finite harmonic morphism $\phi$ of some degree $k$ from a multigraph to a tree, and the existence of a partition tree of breadth $k$ for the associated weighted simple graph. The basic idea is to use the pre-images of 
vertices in $T$ as partitioning sets, but this needs some elaboration.  

\begin{theorem} \label{fhm-to-dt} Suppose $G$ is a weighted simple graph, and $\phi\colon H\to T$ is a finite harmonic morphism  of degree $\deg(\phi) = k$, where is $H$ a loopless refinement of the multigraph corresponding to $G$ and $T$ is a tree. Then one can construct in time $O(k\cdot|V(T)|)$  a tree partition  $\mathcal T = (X,T')$ for a subdivision of $G$ such that \begin{enumerate} 
\item $\width(\mathcal T) \leq k$, and 
\item $|V(T')| \leq 2|V(G)|$.   
\end{enumerate}
In particular, $\mathrm{stb}(G) \leq \sgon(G)$. 
 \end{theorem}

\begin{proof}
Construct a tree partition $\mathcal T = (X, T)$ as follows. For every node $t\in V(T)$, define $X_t = \phi^{-1}(t)\cap V(G)$. For every edge $uv \in E(G)$, do the following. Let $i \in V(T)$ be such that $u\in X_i$ and let $j \in V(T)$ be such that $v\in X_j$. Let $i, t_1, t_2, \ldots, t_l, j$ be the path between $i$ and $j$ in $T$. Subdivide the edge $uv$ into a path $u, s_1, s_2, \ldots, s_l, v$ and add the vertex $s_r$ to the set $X_{t_r}$ for each $r$. Notice that this indeed yields a tree partition. 

We claim that this tree partition has breadth at most $k$. Let $t\in V(T)$, and suppose that $s_i \in X_t$ is a subdivision vertex of an edge $uv \in E(G)$. The edge $uv$ is subdivided in $H$ as well (if $w(uv) > 1$, there are parallel edges $uv$ in the corresponding multigraph, so all such edges are subdivided; pick any), and the path from $u$ to $v$ is mapped by $\phi$ to a walk from $\phi(u)$ to $\phi(v)$. By definition $t$ is in the unique path from $\phi(u)$ to $\phi(v)$, so every walk from $\phi(u)$ to $\phi(v)$ contains $t$. It follows that there is some subdivision vertex in $H$ that is mapped to $t$. We conclude that $|X_t| \leq |\phi^{-1}(t)| \leq k$. The argument for the edges is analogous. We conclude that the breadth of this tree partition is at most $k$. 

Now we will change the tree partition slightly to ensure that the number of nodes is at most $2|V(G)|$. 
First remove all vertices from $T$ for which $X_t = \emptyset$. Notice that the resulting graph $T'$ will still be a tree. Moreover, for every leaf $t$ of $T'$ the set $X_t$ will contain a vertex of $G$. 
For every degree $2$ vertex $t$ of $T'$ for which $X_t$ does not contain a vertex of $V(G)$, contract $t$ with one of its neighbours, and contract all vertices in $X_t$ with a neighbour as well. 
The number of nodes in the resulting tree is at most $2|V(G)|$ since every node either contains a vertex of $G$ or has degree at least 3. 

The runtime analysis is as follows. For every edge $uv \in E(G)$, we can find the (unique) shortest path between $\phi(u)$ and $\phi(v)$ in $T$ in time the length of the path. We can subdivide the edge $uv$ and add the subdivisions to the corresponding sets $X_t$ in time linear in the length of this path. We conclude that the total construction can be done in the sum of the lengths of the paths, which is upper bounded by $k\cdot|V(T)|$ time. 
\end{proof}

\begin{remark}
Theorem \ref{fhm-to-dt} can also be shown using the results of  \cite[Lemma  6.5]{sgon-computable}. The properties of the morphism considered in that paper are stronger than the ones we need here, which allows one to obtain a tree partition with at most $|V(G)|$ nodes. 
\end{remark} 

\begin{example} For the multigraphs in Figure \ref{fig:ex-morphism}, the constructed tree partitions have breadth equal to the stable gonality: for (a), each vertex forms an individual bag, and bags are connected by edges of weight $2$, leading to a total breadth of $2$; for (b), the constructed tree partition consists of one bag containing both non-subdivision vertices, and no edges, again leading to a breadth of $2$. \end{example} 

The main application of the above result is the following reduction for the proof of one part of the main theorem. 

\begin{corollary} 
To prove that a weighted graph problem is $\FPT$ for $\sgon$ \textup{(}given a morphism of a refinement of the corresponding multigraph to a tree whose degree realizes the stable gonality\textup{)}, it suffices to prove that it is $\FPT$ for the breadth of a given tree-partition of a subdivision of the weighted graph.  
\end{corollary} 

\subsection{Problem definitions}
\label{sec:problems}
In this subsection, we give the formal definitions of the flow, orientation, and capacitated problems studied in this paper. We always assume that integers are given in unary. 

We consider five (strongly related) problems that ask for orientations of a weighted undirected graph.

\begin{verse}
{\sc Minimum Maximum Outdegree} (MMO)\\
{\bf Given:} Undirected weighted graph $G=(V,E,w)$ with a weight function $w: E \rightarrow {\Z}_{>0}$; integer $r$\\
{\bf Question:} Is there an orientation of $G$ such that for each $v\in V$, the total weight of all edges directed out of $v$ is at most $r$?
\end{verse}

MMO was shown to be \W-hard for treewidth by Szeider \cite{Szeider11}. 

\begin{verse}
{\sc Circulating Orientation} (CO)\\
{\bf Given:} Undirected weighted graph $G=(V,E,w)$ with a weight function $w: E \rightarrow {\Z}_{>0}$\\
{\bf Question:} Is there an orientation of the edges such that for all $v\in V$, the total weight of all edges directed to $v$ equals the total weight of all edges directed from $v$?
\end{verse}

\begin{verse}
{\sc Outdegree Restricted Orientation} (ORO)\\
{\bf Given:} Undirected weighted graph $G=(V,E,w)$ with a weight function $w: E \rightarrow {\Z}_{>0}$; for each vertex $v\in V$, an 
interval $D_v \subseteq {\Z}_{\geq 0}$\\
{\bf Question:} Is there an orientation of $G$ such that for each $v\in V$, the total weight of all edges directed out of $v$ is an integer in $D_v$?
\end{verse}

\begin{verse}
{\sc Target Outdegree Orientation} (TOO)\\
{\bf Given:} Undirected weighted graph $G=(V,E,w)$ with a weight function $w: E \rightarrow {\Z}_{>0}$; for each vertex $v\in V$, an
integer $d_v$\\
{\bf Question:} Is there an orientation of $G$ such that for each $v\in V$, the total weight of all edges directed out of $v$ equals $d_v$?
\end{verse}

\begin{verse}
{\sc Chosen Maximum Outdegree} (CMO)\\
{\bf Given:} Undirected weighted graph $G=(V,E,w)$ with a weight function $w: E \rightarrow {\Z}_{>0}$; for each vertex $v\in V$, an
integer $m_v$\\
{\bf Question:} Is there an orientation of $G$ such that for each $v\in V$, the total weight of all edges directed out of $v$ is at most $m_v$?
\end{verse}

We also consider two (variants of) classical graph flow problems, the first of which is couched using orientations.

\begin{verse}
 {\sc Undirected Flow with Lower Bounds} (UFLB) \\
 {\bf Given:} Undirected graph $G=(V,E)$, for each edge $e\in E$ a positive integer capacity $c(e) \in \Z_{>0}$ and a non-negative integer lower bound $\ell(e) \in \Z_{\geq 0}$, vertices $s$ (source) and $t$ (target), a non-negative integer $R \in \Z_{>0}$ (value) \\
 {\bf Question:} Is there an orientation of $G$ such that the resulting directed graph $D$ allows an $s$-$t$-flow $f$ that meets capacities and lower bounds (i.e., $\ell(a) \leq f(a) \leq c(a)$ for all arcs in $D$), with value $R$? 
\end{verse}

ULFB was shown to be strongly NP-complete by Itai \cite[Theorem 4.1]{Itai}.

\begin{verse}
{\sc All-or-Nothing Flow} (AoNF) \\
{\bf Given:} Directed graph $G=(V,E)$, for each arc a positive capacity $c(e)$, vertices $s$, $t$, positive integer $R$\\
{\bf Question:} Is there a flow $f$ from $s$ to $t$ with value $R$ such that for each arc $e\in E$,
$f(e)=0$ or $f(e)=c(e)$?
\end{verse}

An NP-completeness proof of {\sc All or Nothing Flow} is given in \cite{Alexandersson01}.

Finally, we consider a coloured and uncoloured capacitated version of \textsc{Dominating Set}. 

\begin{verse}
 {\sc Capacitated Dominating Set} (CDS) \\
 {\bf Given:} Undirected graph $G=(V,E)$, for each vertex $v\in V$ a positive integer capacity $c(v) \in \Z_{>0}$, integer $k$ \\
 {\bf Question:} Is there a set $D \subset V$ of size $|D| \leq k$ and a function $f \colon V \setminus D \rightarrow D$ such that $vf(v) \in E$ for all $v\in V\setminus D$ and $|f^{-1}(v)| \leq c(v)$ for all $v \in D$?   
\end{verse}

CDS was shown to be \W-hard for treewidth in \cite{Dom}, even when restricted to planar graphs \cite{Bod-plain}. Recently, it was shown that
the problem is XNLP-complete for pathwidth~\cite[Theorem 8]{BodlaenderGJ22}. The usefulness of the following auxiliary coloured version of the problem was first pointed out in \cite{Fomincolor}.  

\begin{verse}
 {\sc Capacitated Red-Blue Dominating Set} (CRBDS) \\
 {\bf Given:} Undirected bipartite graph $G=(V=R \sqcup B,E)$, for each ``red'' vertex $v\in R$ a positive integer capacity $c(v) \in \Z_{>0}$, integer $k$ \\
 {\bf Question:} Is there a set $D \subset R$ of size $|D| \leq k$ and a function $f \colon B \rightarrow D$ such that $bf(b) \in E$ and $|f^{-1}(v)| \leq c(v)$ for all $v \in D$?   
\end{verse}

\section{Transformations between orientation and flow problems}
\label{section:transformations}

\begin{figure}
\centering
\begin{subfigure}{.3\textwidth}
\begin{tikzpicture}[scale=.75]
\node (TOO) at (0,0) {TOO};
\node (AoNF) at (-2,0) {AoNF};
\node (ORO) at (-2, -2) {\textbf{ORO}};
\node (CMO) at (0, -2) {CMO};
\node (MMO) at (0, -4) {MMO};
\node (CO) at (2, -2) {CO};
\node (UFLB) at (2, -4) {UFLB};
\draw[->] (TOO) -- (ORO);
\draw[->] (CMO) -- (ORO);
\draw[->] (MMO) -- (CMO);
\draw[->] (CO) -- (TOO);
\draw[->] (UFLB) -- (CO);
\draw[->] (AoNF) -- (TOO);
\end{tikzpicture}
\caption{}
\end{subfigure}
\quad
\begin{subfigure}{.3\textwidth}
\begin{tikzpicture}[scale=.75]
\node (TOO) at (0,0) {TOO};
\node (AoNF) at (-2,0) {\textbf{AoNF}};
\node (ORO) at (-2, -2) {ORO};
\node (CMO) at (0, -2) {CMO};
\node (MMO) at (0, -4) {MMO};
\node (CO) at (2, -2) {CO};
\node (UFLB) at (2, -4) {UFLB};
\draw[->] (AoNF) -- (TOO);
\draw[->] (TOO) -- (ORO);
\draw[->] (TOO) -- (CMO);
\draw[->] (CMO) -- (MMO);
\draw[->] (TOO) -- (CO);
\draw[->] (CO) -- (UFLB);
\end{tikzpicture}
\caption{}
\end{subfigure}
\quad
\begin{subfigure}{.3\textwidth}
\begin{tikzpicture}[scale=.75]
\node (TOO) at (0,0)  {\textbf{TOO}};
\node (ORO) at (-2, -2) {ORO};
\node (CMO) at (0, -2) {CMO};
\node (MMO) at (0, -4) {MMO};
\node (CO) at (2, -2) {CO};
\node (UFLB) at (2, -4) {UFLB};
\draw[->] (TOO) -- (CMO);
\draw[->] (CMO) -- (MMO);
\draw[->] (CMO) -- (ORO);
\draw[->] (TOO) -- (CO);
\draw[->] (CO) -- (UFLB);
\end{tikzpicture}
\caption{}
\end{subfigure}
\caption{Transformation between different problems with respect to parameter (a) treebreadth or stable gonality, for which ORO is FPT, (b) pathwidth, for which AoNF is \XNLP-complete, and (c) vertex cover number, for which TOO is \W-hard. }
\label{fig:transformations}
\end{figure}
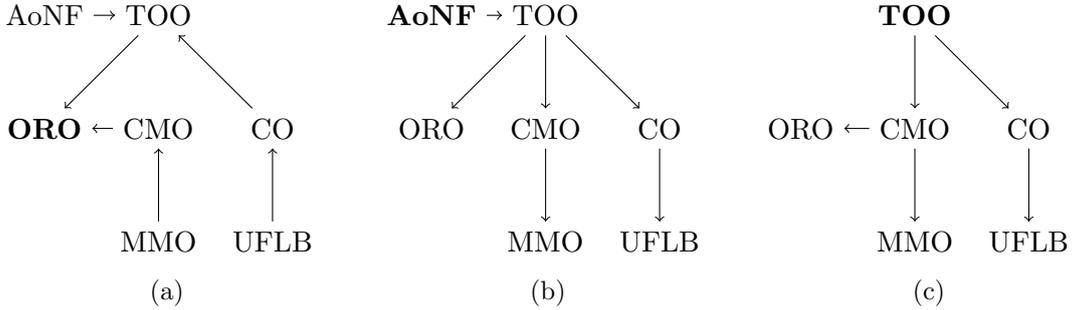

In this section, we give a number of relatively simple transformations between seven of the main problems that
we study in this paper. A summary of the transformations can be found in Figure \ref{fig:transformations}. In Section \ref{section:algorithm}, we show that ORO is fixed parameter tractable when parameterized by the stable gonality. In Section \ref{sec:hardness}, we show that AoNF is XNLP-hard when parametrized by pathwidth and that TOO is $W[1]$-hard when parameterized by the vertex cover number. The algorithms and hardness results for the other problems follow from these reductions. 

First, a number of these problems can be seen as special cases of others:
\begin{itemize}
    \item {\sc Target Outdegree Orientation} is the special case of {\sc Outdegree Restricted Orientation}
    by taking each interval $D_v = \{d_v\} = [d_v,d_v]$ a singleton.
    \item {\sc Chosen Maximum Outdegree} is the special case of {\sc Outdegree Restricted Orientation}
    by taking each interval $D_v = [0,m_v]$ starting at $0$.
    \item {\sc Minimum Maximum Outdegree} is the special case of {\sc Chosen Maximum Outdegree} where
    all values $m_v$ are equal to $r$.
    \item {\sc Circulating Orientation} is the special case of {\sc Target Outdegree Orientation} where
    for each vertex $v$, we set $d_v = \frac{1}{2}\deg(v)$, with $\deg(v)$ equal to the sum of all weights of edges incident to $v$.
    \item {\sc Circulating Orientation} is the special case of {\sc Undirected Flow with Lower Bounds} where we set for each edge the lower bound equal to its capacity, and the target 
    value of the flow to $0$.
\end{itemize}

\subsection{From \textmd{\textsc{All-Or-Nothing Flow}} to \textmd{\textsc{Target Outdegree Orientation}}}
We now show how to transform instances of {\sc All-Or-Nothing Flow} to equivalent instances of {\sc Target Outdegree Orientation}, see Lemma~\ref{lemma:aonf2too}. The transformation increases the pathwidth
of the graph by at most one, and does not change the stable gonality of the graph.

This transformation has two consequences. On the positive side,
when we have bounded stable gonality,
we can use the algorithm for {\sc Outdegree Restrictred Orientation} from Section~\ref{section:algorithm};
on the negative side, combined with Theorem~\ref{theorem:allornothingxnlp}, this shows that
{\sc Target Outdegree Orientation} with pathwidth as parameter is XNLP-hard.

\smallskip

Suppose we are given
an instance ($G$, $s$, $t$, $c$, $R$) of {\sc All-Or-Nothing Flow}. 
The transformation uses a number of steps. 

First, 
we subdivide all arcs, with both subdivisions having the same capacity. Note that this step increases the
pathwidth by at most one, and does not increase the stable gonality. 
(However, the step can increase the size of a minimum vertex cover.) Also note that this
creates an equivalent instance of {\sc All-or-Nothing Flow}.

We call the resulting graph $G^1$, and denote the capacity function again by $c$.

In the second step, we create an equivalent instance of {\sc Target Outdegree Orientation} except that
we allow weights of edges to be a multiple of $1/2$. 

Let $G^2=(V^2,E^2)$ be the underlying undirected graph of $G^1$; i.e., $G^2$ is obtained by dropping all
directions of arcs in $G^2$. Note that $G^2$ is a simple graph, as a result of the step where we
subdivided all arcs.

If $uv$ is an arc in $G^1$ with capacity $c(uv)$, then the weight of the undirected edge
$uv$ in $G^2$ is $w(uv) = c(uv)/2$. 

Let $\text{IN}_{G^2}(v)$
be the set of edges in $G^2$ whose originating arc was directed towards $v$,
i.e., all edges $uv$ in $G^2$ with $uv$ an arc in $G^1$.
Let $\text{OUT}_{G^2}(v)$ be the set of edges in $G^2$ whose originating arc was directed out of $v$,
i.e., all edges $uv$ in $G^2$ with $vu$ an arc in $G^1$.

For each vertex $v\in V^2$, we compute the target outdegree value $d_v$, as follows.
\begin{itemize}
    \item Set $\beta_v$ to be the sum of the capacities of all arcs that are directed towards $v$ in $G^1$.
    \item Now, if $v\in V \setminus \{s.t\}$, then set $d_v = \beta_v/2$.
    Set $d_s = R/2 + \beta_s/2$, and $d_t = -R/2 + \beta_t/2$.
\end{itemize}

\begin{lemma}
There is an all-or-nothing flow in $G^1$ from $s$ to $t$ with value $R$, if and only there is an orientation of $G^2$
such that for each vertex $v$, the total weight of all outgoing edges of $v$ 
is equal to $d_v$.
\end{lemma}

\begin{proof}
Suppose that we have an all-or-nothing flow $f$ in $G^1$ from $s$ to $t$ with value $R$. Consider an
arc $xy$ in $G^1$. If $f$ sends a positive amount of flow over this arc (and thus, flow equal to the
capacity),
then orient the edge $xy$ from $x$ to $y$. Otherwise, $0$ flow is send over the arc, and we
orient the edge $xy$ from $y$ to $x$, i.e., edges are oriented in the same direction as their corresponding arc when they have non-zero flow, and in the opposite direction as their corresponding arc when the
flow over the arc is $0$. (The step is inspired by the principle of cancelling flow.)

We now verify that this orientation meets the targets. First, consider a vertex $v\in V\setminus \{s,t\}$.
Suppose the total inflow to $v$ by $f$ is $\delta_v$. Then, the total outflow from $v$ by $f$ is also
$\delta_v$, due to flow conservation.
We start by looking at the edges from $\text{IN}_{G^2}(v)$. The total capacity of all arcs that send flow to $v$ is
$\delta_v$, so the total weight of all edges from $\text{IN}_{G^2}(v)$ that are directed towards $v$ equals
$\delta_v/2$. The total weight of all edges from $\text{IN}_{G^2}(v)$ equals $\beta_v/2$.
So, the total weight of all edges from $\text{IN}_{G^2}(v)$ that are directed out of $v$ equals
$(\beta_v - \delta_v)/2$.
Now, look at the edges in $\text{OUT}_{G^2}(v)$. The total capacity of all arcs that send flow out from $v$ is
again $\delta_v$, so the total weight of all edges in $\text{OUT}_{G^2}(v)$ that are directed out of $v$ 
equals $\delta_v/2$.
Thus, the total weight of the edges that are directed out of $v$ in the orientation equals 
$(\beta_v - \delta_v)/2 + \delta_v/2 = \beta_v/2$.

Now, consider $s$. Suppose $f$ has inflow $\delta_s$ at $s$, and thus outflow $R+\delta_s$. Again,
the total weight of edges from $\text{IN}_{G^2}(s)$ that are directed out of $s$ equals $(\beta_s-\delta_s)/2$. The total weight of the edges from $\text{OUT}_{G^2}(s)$ equals $(R+\delta_s)/2$.
So, the total weight of all outgoing arcs in the orientation is $R/2 + \beta_s/2$. The argument for $t$ is similar.

Suppose that we have an orientation of $G^2$
such that each vertex has total weight of all edges directed out of $v$ equal to $d_v$.
If an edge is oriented in the same direction as its originating arc, then send
flow over the arc equal to its capacity, otherwise,
send $0$ flow over that arc. Let $f$ be the corresponding function.
We claim that $f$ is a flow in $G^1$ from $s$ to $t$ with value $R/2$.

Suppose the inflow by $f$ to a vertex $v\in V$ equals $\epsilon_v$. 
The total weight of edges in $\text{IN}_{G^2}(v)$ that are directed towards $v$ thus is $\epsilon_v/2$.
The total weight of all edges in $\text{IN}_{G^2}(v)$ equals $\beta_v/2$.
Hence, the total weight of all edges in $\text{IN}_{G^2}(v)$ that are oriented out of $v$ is
$\beta_v/2 - \epsilon_v/2$. Hence, the total weight of all edges in $\text{OUT}_{G^2}(v)$ that are oriented 
out of $v$ is $d_v - (\beta_v/2 - \epsilon_v/2)$. This value equals
$\epsilon_v/2$ for $v\not\in\{s,t\}$,  $R/2 + \epsilon_v/2$ for $v=s$, and
$-R/2 + \epsilon_v/2$ for $v=t$. The total amount of flow send out of $v$ by $f$ is twice this number,
so for $v\not\in \{s,t\}$, the inflow equals the outflow, namely $\epsilon_v$; for
$v=s$, the outflow is $R$ larger than its inflow, and for $v=t$, the outflow is $R$ smaller than its inflow.
The result follows.
\end{proof}

We now have shown that we can transform an instance of {\sc All-or-Nothing Flow} to an equivalent 
instance of {\sc Target Outdegree Orientation}, except that in the latter problem, values are
allowed to be a multiple of $\frac{1}{2}$. To obtain an instance of {\sc Target Outdegree Orientation}
with all values integral, we multiply all weights and outdegree targets by two.

Note that the undirected graph $G^2$ is obtained from the directed graph $G$ by subdividing each
edge, and then dropping directions of edges. As above, the operation increases the pathwidth by at most one, and 
does not increase the (stable) treebreadth and stable gonality, i.e.\ sgon of the undirected graph underlying $G$ equals sgon of $G^2$. One easily observes that the transformations can be carried
out in polynomial time and logarithmic space. The latter is needed for XNLP-hardness 
proofs.

\begin{lemma}
There is a parameterized log-space reduction from {\sc All-or-Nothing Flow} 
to {\sc Target Outdegree Orientation} with respect to parameters pathwidth, (stable) treebreadth and   stable gonality, which also transforms the associated given finite harmonic morphism from a refinement of the input of degree $\sgon(G)$ to one of the new
graph with the same degree or transforms the given tree partition of (a refinement of) $G$ of breadth $\textup{(s)tb}(G)$ to one of the new graph with the same breadth.
\label{lemma:aonf2too}
\end{lemma}

\subsection{From  {\sc Chosen Maximum Outdegree} to {\sc Minimum Maximum Outdegree}}
Szeider~\cite{Szeider11} gives a transformation from {\sc Chosen Maximum Outdegree} to {\sc Minimum Maximum Outdegree}.
This reduction changes the graph in the following way: two additional vertices are added, as well as a number of edges --- each new edge has at least one of the two new vertices as endpoint.
Thus, the vertex cover number and the pathwidth of the graph is increased by at most two by this reduction.

\begin{lemma}[Szeider~\cite{Szeider11}]
There is a parameterized log-space reduction from \textsc{Chosen Maximum Outdegree} to \textsc{Minimum Maximum Outdegree} for the parameterizations by pathwidth and by vertex cover number.
\label{lemma:cmo2mmo}
\end{lemma}

\subsection{From \textmd{\textsc{Target Outdegree Orientation}} to \textmd{\textsc{Circulation Orientation}}}
Again, the following reduction is inspired by standard insights from network flow theory.

\begin{lemma}
There is a parameterized log-space reduction from \textsc{Target Outdegree Orientation} to \textsc{Circulation Orientation} when parameterized by pathwidth or by vertex cover number.
\label{lemma:too2co}
\end{lemma}

\begin{proof}
Let $(G, w, d_v)$ be an instance of \textsc{Target Outdegree Orientation}. We turn the instance into a circulation, and do this with a construction that is classical in flow theory, see e.g. \cite{flow}. 
First we determine the demand of each vertex. If a vertex $v$ has outdegree bound $d_v$, and total weight of incident edges $r_v$, then the indegree will be $r_v-d_v$, and this gives in a flow a demand of $r_v-2d_v$. Let $\alpha$ be the sum of all positive demands, that is, $\alpha = \frac{1}{2} \sum_{v\in V(G)} |r-2d_v|$. 
The construction is to add a supersource $s$, a supersink $t$, and an edge from $t$ to $s$ with weight $\alpha$. Moreover, we add an edge with weight $r_v-2d_v$ from $s$ to each vertex $v$ with negative demand, and an edge with weight $r_v-2d_v$ from each vertex $v$ with positive demand to $t$.
Call $H$ the thus constructed graph.

\begin{claim}
If $G$ is a yes-instance of \textsc{Target Outdegree Orientation}, then $H$ is a yes-instance of \textsc{Circulating Orientation}. 
\end{claim}
\begin{claimproof}
Suppose that there is an orientation of $G$ such that each vertex $v\in V(G)$ has weighted outdegree $d_v$. Then we can extend this orientation to a circulation of $H$ by directing all edges $vt$ from $v$ to $t$, all edges $sv$ from $s$ to $v$ and the edge $st$ from $t$ to $s$. 

To show that this is indeed an orientation, we distinguish cases. For each vertex $v$ with $r_v-2d_v = 0$, the outdegree is $d_v$ and the indegree is $r_v - d_v = d_v$. For each vertex $v$ with $r_v-2d_v > 0$, the outdegree is $d_v + r_v - 2d_v = r_v - d_v$, since the edge $vt$ is oriented out of $v$, and the indegree is $r_v - d_v$, which indeed equals the outdegree. The case when $r_v-2d_v <0$ is similar. The in- and outdegree of $s$ and $t$ is $\alpha$. 
\end{claimproof}

\begin{claim}
If $H$ is a yes-instance of \textsc{Circulating Orientation}, then $G$ is a yes-instance of \textsc{Target Outdegree Orientation}. 
\end{claim}
\begin{claimproof}
Suppose that there is an orientation of $H$ such that each vertex $v\in V(H)$ has equal weighted in- and outdegree. Suppose that the edge $st$ is oriented from $t$ to $s$. Then $t$ has outdegree $\alpha$, so all edges $vt$ are oriented towards $t$. Now consider a vertex $v$ with $r_v-2d_v > 0$. It follows that $\indeg_G(v) = \outdeg_G(v) + r_v-2d_v$, where $\indeg_G(v)$ and $\outdeg_G(v)$ are the in- and outdegree of $v$ restricted to the edges of $G$. Since $\indeg_G(v) = r_v - \outdeg_G(v)$, it follows that $\outdeg_G(v) = d_v$. The case of vertices with $r_v - 2d_v \leq 0$ is similar. 
We conclude that restricting the orientation to $G$ will give the desired outdegree for all vertices. 

If the edge $st$ is oriented from $s$ to $t$, then flipping the orientation of all edges gives another circulating orientation, and the result follows as above. 
\end{claimproof}

Notice that this is a polynomial time and logarithmic space construction. Moreover, the pathwidth of $H$ is at most $\tw(G) + 2$, which can be seen by adding $s$ and $t$ to all bags of a path decomposition of $G$. 
Also, the vertex cover number of $H$ is at most two more than the vertex cover number of $G$: if $S$ is a vertex cover of $G$, then $S\cup\{s,t\}$ is a vertex cover of $H$.
\end{proof}

\subsection{From \textmd{\textsc{Target Outdegree Orientation}} to \textmd{\textsc{Chosen Maximum Outdegree}}}
With a pigeonhole argument, we obtain a simple reduction from \textsc{Target Outdegree Orientation} to \textsc{Chosen Maximum Outdegree}.

Consider an instance of \textsc{Target Outdegree Orientation}, i.e., we are given an undirected
graph $G=(V,E)$, for each edge $e\in E$ a positive integer weight $w(e)$, and for each vertex
$v\in V$ a positive integer target value $d_v$. We have the following two simple observations.

\begin{lemma}
If $G$ has an orientation such that for each vertex $v\in V$ the sum of the weights of edges directed
out of $v$ equal to $d_v$, then $\sum_{e\in E} w(e) = \sum_{v\in V} d_v $.
\end{lemma}

\begin{proof}
For a given orientation, each edge is directed out of exactly one vertex.
\end{proof}

\begin{lemma}
Suppose $\sum_{e\in E} w(e) = \sum_{v\in V} d_v $. For each orientation of $G$, we have that
for each vertex $v\in V$ the total weight of edges directed out of $v$ equals $d_v$, if and only if
for each vertex $v\in V$ the total weight of edges directed out of $v$ is at most $d_v$.
\end{lemma}

\begin{proof}
Consider an orientation of $G$, and suppose that for each vertex $v\in V$ the total weight of edges directed out of $v$ is at most $d_v$. If there is a vertex $u$ for which the total weight of edges
directed out of $u$ less than $d_u$, then the sum over all vertices $v\in V$ of the total weight 
of the edges directed out of $V$ is less than $\sum_{v\in V} d_v$. But each edge $e\in E$ is counted
once in this sum, hence $\sum_{v\in V} d_v < \sum_{e\in E} w(e)$, a contradiction.

The other direction is trivial.
\end{proof}

\begin{lemma}
There is a parameterized log-space reduction from \textsc{Target Outdegree Orientation} to \textsc{Chosen Maximum Outdegree}  parameterized by pathwidth, vertex cover number, or stable gonality.
\label{lemma:too2cmo}
\end{lemma}

\begin{proof}
The lemmas above show that we can use the following reduction: first, check whether $\sum_{e\in E} w(e) = \sum_{v\in V} d_v $. If not,
then reject (or transform to a trivial no-instance); otherwise, set $m_v=d_v$ for each $v$.
As $G$ is not changed, pathwidth, vertex cover number and stable gonality are the same.
\end{proof}

\subsection{From \textmd{\textsc{Undirected Flow with Lower Bounds}} to {\sc Circulating Orientation}}

\begin{lemma}\label{lemma:uflb2co-sgon}
Suppose we have an instance of \textsc{Undirected Flow with Lower Bounds} with a tree partition
 of breadth at most $k$. Then we can build, in polynomial time and logarithmic space, an equivalent instance of
{\sc Circulating Orientation} with a tree partition of breadth $O(k^2)$ .
\end{lemma}

\begin{proof}
The transformation is done in four steps.

Suppose we are given an instance of {\sc Undirected Flow with Lower Bounds}, i.e., an
undirected graph $G=(V,E)$, capacity function $c: E \rightarrow \Z_{>0}$, lower bound function
$\ell: E \rightarrow \Z_{\geq 0}$, vertices $s, t\in V$, and target flow value $R\in \Z_{\geq 0}$.

First, we turn the instance into an orientation problem, or, equivalently,
an instance of {\sc Undirected Flow with Lower Bounds} with flow value $0$ (a circulation).
This is done by adding an edge $st$ with lower bound and capacity equal to $R$.
Let $G^1$ be the resulting graph. (This step is skipped when $R=0$.)

\begin{claim}
There is a flow in an orientation of $G$ from $s$ to $t$ with value $R$ fulfilling lower bounds,
if and only if there is a flow in an orientation of $G^1$ from $s$ to $t$ with value $0$ fulfilling lower bounds.
\end{claim}

\begin{claimproof}
Suppose there is a flow $f$ in an orientation of $G$ from $s$ to $t$ with value $R$ fulfilling lower bounds.
Orient the new edge from $t$ to $s$ and send $R$ flow over this edge. All other edges have their flow
dictated by $f$. This gives the desired solution for $G^1$.

Suppose there is a flow in an orientation of $G^1$ from $s$ to $t$ with value $0$ fulfilling lower bounds.
If the edge $st$ is oriented from $s$ to $t$, then reverse the direction of all arcs in the orientation,
but send the same amount of flow over each edge (but now in the opposite direction). This gives an
equivalent solution. We can now assume that the edge $st$ is oriented from $t$ to $s$. Deleting this
edge and its flow gives the desired orientation and flow for $G$.
\end{claimproof}

We now assume that $R=0$, and thus, the flow we look for is a circulation.
The vertices $s$ and $t$ no longer play
a special role as flow conservation also holds for $s$ and $t$.

\begin{figure}
    \centering
    \includegraphics[width=0.7\textwidth]{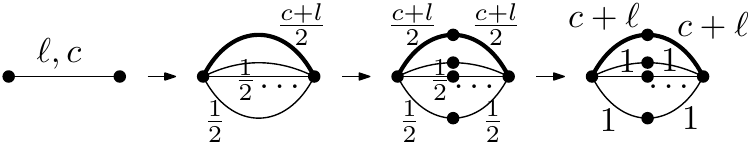}
    \caption{The second, third and fourth step of the transformation for a single edge}
    \label{figure:transformundirectedlow}
\end{figure}

The second, third and fourth step are illustrated in Figure~\ref{figure:transformundirectedlow}, where the
steps that are applied to a single edge are shown.

In the second step, we create an intermediate undirected graph with parallel edges. These edges
have weights that are a multiple of $1/2$. 
This is done as follows. Suppose we have an edge $e$ with capacity $c(e)>0$ and lower bound
$\ell(e)\geq 0$; $0 \leq \ell(e) \leq c(e) $. Now, replace $e$ by the following parallel edges: one
{\em heavy}
edge of weight $(c(e)+\ell(e))/{2}$, 
and $c(e)-\ell(e)$ {\em light} edges of weight $1/2$. 
Let $G^2$ be the resulting multigraph. 

\begin{claim} 
There is an integer flow with value 0 respecting lower bounds and capacities in an orientation of $G^1$, if and only if there
is a circulating orientation in $G^2$.
\end{claim}

\begin{claimproof}
Suppose $f$ is an integer flow with value 0, or equivalently, a circulation, that respects lower bounds and capacities in an orientation of $G^1$.  

For each
edge $uv\in E$, suppose $f$ send $\alpha$ units of flow from $u$ to $v$ in $G^1$. Now, in $G^2$, we
orient the heavy edge (with weight $(c(e)+\ell(e))/{2}$) from $u$ to $v$. 
Orient $(\alpha - \ell(e))$ of the light parallel edges (of weight $1/2$ between $u$ and $v$ from $u$ to $v$ and all other of these light parallel edges
from $v$ to $u$. Thus, we have $c(e)-\ell(e) - (\alpha - \ell(e)) = c(e) - \alpha$ light edges
directed from $v$ to $u$.

Thus, the heavy edge sends $(c(e)+\ell(e))/{2}$ flow from $u$ to $v$; the light edges
send $(\alpha-\ell(e))/{2}$ flow from $u$ to $v$ and
$(c(e)-\alpha)/{2}$ flow from $v$ to $u$. The net flow contribution that goes from $u$ to $v$ of
all these edges adds up to 
\[ \frac{c(e)+\ell(e)}{2} + \frac{\alpha-\ell(e)}{2} - \frac{c(e)-\alpha}{2} = \alpha.
\]

Thus, the flow directed by the constructed orientation of the multigraph is for each pair of
adjacent vertices the same as the flow in $f$. As the latter is a circulation, the constructed
orientation of the multigraph is also an orientation.

Suppose we have an orientation that gives a circulation in $G^2$. Consider an edge $uv\in E$.
Suppose the `heavy' edge between $u$ and $v$ (with weight $(c(e)+\ell(e))/{2}$) is
oriented from $u$ to $v$. Then, in $G^1$, we send flow in the direction from $u$ to $v$.
Suppose $\gamma$ light edges of weight $1/2$ are oriented from $u$ to $v$;
thus $c(e)-\ell(e) - \gamma$ light
edges (of weight $1/2$) are oriented from $v$ to $u$.
Now, send \[\frac{c(e)+\ell(e)}{2} + \gamma \cdot \frac{1}{2} - \left(c(e)-\ell(e) - \gamma\right) \cdot \frac{1}{2}
= \ell(e) + \gamma\]
flow from $u$ to $v$ over the edge $uv$ in $G^1$. Thus, the flow from $u$ to $v$ in $G^1$ equals the net flow
from $u$ to $v$ over all parallel edges in $G^2$ (where flow from $v$ to $u$ cancels the same amount of
flow from $u$ to $v$, as usual in flow theory). 
Note that the flow sent from $u$ to $v$ is an integer 
in $[\ell(e),c(e)]$. As the difference of what is sent from $u$ to $v$ and what is sent from
$v$ to $u$ in $G^2$ equals the amount sent in $G^1$, flow conservation also holds for the flow in $G^1$,
and we again have a circulation.
\end{claimproof}

The last two steps are relatively simple.
In the third step, we turn the graph into a simple graph (without parallel edges), by subdividing each edge.
When subdividing an edge, the two resulting edges get the same lower bound and capacity as the original  edge. One easily sees that this step gives equivalent instances.

In the fourth step, we obtain an equivalent instance with only integral values by multiplying all
capacities and lower bounds by two. Let $G^4$ be the resulting graph.

Using these four steps, we transformed an instance of {\sc Undirected Flow with Lower Bounds} into an
equivalent instance of {\sc Circulating Orientation}. 

Now, if we have a tree partition of $G$ of breadth $k$, we can build a tree partition of a subdivision
of $G^4$ of breadth at most $O(k)$ as follows. We first build a tree partition of $G^1$. When $s$ and $t$ are 
in the same or adjacent bags, then we do not need to change the graph or tree partition.
When $s$ and $t$
are not in the same or adjacent bags, then suppose the path in $\cal T$ from the bag containing $s$ to the bag
containing $t$ has $q$ intermediate nodes. Subdivide the edge $st$ $q$ times, and place in each intermediate
node of this path between the bags one of the subdivision nodes, in order. This increases the breadth of
the tree partition by at most $R$. Now, note that if $s$ and $t$ are not in the same bag, then take the edges between
the bag containing $s$ and the neighbouring bag on the path in $\cal T$ towards the bag containing $t$. These
form an $s$-$t$ cut of size at most the breadth of the tree partition. Hence, if $R>k$, we can reject (by the
minimum cut maximum flow theorem.) So, the step increases the breadth by at most $k$.

We need to change the tree partition again when we subdivide the graph in the third step.
First, consider edges between vertices in different bags. To accommodate the subdivisions of these edges, we subdivide each arc $ij$ of $\cal T$ and place the subdivisions of edges between a vertex in $X_i$ and a vertex in $X_j$ in this new bag. Each vertex in this new bag can be associated with an edge of weight at least one with one endpoints in $X_i$ and $X_j$, thus these new bags have $O(k)$ vertices. 
Second, we may subdivide the edge $st$ one or more additional times, such that each bag on the path from $s$ to $t$ contains exactly one subdivision vertex of the edge $st$.
Third, for each adjacent pair of vertices $v,w\in X_i$ in the same bag, 
we add the subdivision vertex of the heavy edge between $v$ and $w$ in $X_i$,
and for each light edge, we add an additional bag that is made incident to $i$.
As there are $O(k^2)$ pairs of vertices in a bag, this steps increases the breadth 
by $O(k^2)$.
The result is a tree partition of the graph obtained in the third step, of
breadth $O(k^2)$. 

The fourth step does not change the graph. Doubling the weight of the edges can double the breadth of the tree partition. Thus the result is a tree partition of breadth $O(k^2)$.
\end{proof}

\begin{remark}
The above reduction is also a reduction with respect to the parameter $\sgon$. 
Let $(G, c, l, R)$ be an instance of \textsc{Undirected Flow with Lower Bounds} and $\phi\colon G' \to T$ a finite harmonic morphism of degree $k$. Recall that $G'$ is a refinement of the multigraph corresponding to $G$. We obtain a morphism $\phi'\colon {G^{4\prime}} \to T'$ with degree $O(4k)$ as follows. If $\phi(s) = \phi(t)$ subdivide the new edges from $s$ to $t$ once, and map the new vertices to unique new leaves. Otherwise, subdivide the edges from $s$ to $t$  into $l$ edges, where $l$ is the length of the $\phi(s),\phi(t)$-path, and map those new vertices to the $\phi(s),\phi(t)$-path. Refine the graph such that the morphism becomes harmonic again. This results in a morphism of degree $O(2k)$, as above. In step 3, when subdividing all edges, subdivide all edges of $T$ as well. This does not change the degree of the morphism. Finally, multiplying all capacities by two means doubling all edges in the corresponding multigraph, and this results in a morphism of degree $O(4k)$. 
\end{remark}

\section{An algorithm for {\sc Outdegree Restricted Orientation} for graphs with small (stable) treebreadth}
\label{section:algorithm}
In this section, we give our main result, and show that {\sc Outdegree Restricted Orientation} is fixed parameter
tractable for graphs with small stable treebreadth, given a tree partition of a subdivision graph $G'$ realizing the breadth. 

\paragraph{Structure of the algorithm}

We give each edge in $G'$ the same weight as it has in $G$; if an edge $e'$ 
resulted from subdividing edge $e$, then its weight is set to $w(e)$.
If vertex $x_e$ resulted from subdividing edge $e$, then we set $D_{x_e}=[w(e),w(e)]$.

With this setting of weights and targed intervals, the problem on $G'$ is equivalent
to the problem on $G$. From this, it follows that we can assume that we have
a tree partition of the input graph itself, of given breadth.

\begin{claim}
There is an orientation of the edges in $G$ with for each $v\in V$, the total weight of all edges directed
out of $v$ in $D_v$, if and only if there is such an orientation in $G'$.
\end{claim}

\begin{claimproof}
If we have the desired orientation in $G$, then orient each edge created by a subdivision in the same way as the original one. If we have the desired orientation in $G'$, then note that for each vertex created by
a subdivision, one of its incident edges is incoming and one is outgoing. Thus, we can orient the original edge in the direction that all its subdivisions use.
\end{claimproof}

Next, we add a new root vertex $r$ to the tree partition, and set $X_r = \emptyset$.

After these preliminary steps, we perform a dynamic programming algorithm on the resulting tree partition $\cal T$, as follows.

For each arc $a$ from a node to its parent, we compute a table $A_a$. We do this bottom-up in $\cal T$.
If the table of the arc to the root has a positive entry, then accept, otherwise reject. Correctness
of this follows from the definition of the information in the tables, as will be discussed below.

\paragraph{Notation}
Note that we assume we are given a graph $G'$ with a tree partition $\cal T$
of breadth at most $k$. We denote the vertices in $G'$ by $V$, the edges by $E$, 
the weight function by $w$, and for each vertex $v$ its target interval by $D_v$.

An orientation of the edges in $G'$ is said to be {\em good}, 
if for each vertex $v\in V$, the total
weight of edges directed out of $v$ is an element of $D_v$.

For a node $i\in I$, we denote the union of all vertex sets $X_j$ with $j=i$ or $j$ a descendant of $i$ 
as $V_i$. 

For an arc $a = i i'\in F$
with $i$ the parent of $i'$, we write $E_a$ for the set of all edges of $G'$
with one endpoint in $V_i$ and one endpoint in $V_{i'}$. I.e., we take all edges with both
endpoints in $V_i$ except the edges with both endpoints in $X_i$: $E_a = (V_i \times V_i \cap E) \setminus X_i \times X_i$.

A {\em partial solution} for arc $a$ is an orientation of $E_a$ such that for each vertex $v\in V_{i'}$,
the total weight of all edges directed out of $v$ is an integer in $D_v$.
Note that for partial solutions, the condition is not enforced for vertices in $X_i$.

Let $\rho$ be a partial solution for $a$. The {\em fingerprint} of $\rho$ is the
function $f: X_i \rightarrow {\Z_{\geq0}}$ where for each $v\in X_i$, $f(v)$ equals the total weight of all
edges directed out of $v$ for the orientation $\rho$.

We say that a partial solution $\rho$ for $a$ is {\em extendable}, if there is a good orientation $\rho'$ of $G'$ with
all edges in $E_a$ oriented in the same way in $\rho$ and $\rho'$.

\paragraph{Some observations}

\begin{claim}
$G'$ has a good orientation if and only if there is a partial solution for the arc between the root and its child.
\end{claim}

\begin{claimproof}
Let $i$ be the unique child of root $r$. Observe that $E_{ri} = E$ and $V_{i} = V$, and thus, a partial solution
for $ri$ is a good orientation, and vice versa.
\end{claimproof}

\begin{claim}
Let $f$ be the fingerprint of a partial solution for $a=ii'$. For all $v\in X_i$, $0\leq f(v)\leq k$.
\end{claim}

\begin{claimproof}
All edges in $E_a$ with $v\in X_i$ as endpoint have their other endpoint in $X_{i'}$, so use the
arc $a$, and thus the total weight of all such edges is at most $k$. 
\end{claimproof}

\begin{claim}
Let $\rho_1$ and $\rho_2$ be partial solutions for $a$ with the same fingerprint. Then $\rho_1$ is
extendable if and only if $\rho_2$ is extendable.
\end{claim}

\begin{claimproof}
Suppose $\rho$ extends $\rho_1$. Consider the orientation that orients all edges in $E_a$ as in $\rho_2$
and all edges in $E\setminus E_a$ as in $\rho$. One easily checks that this is an extension of $\rho_2$.
\end{claimproof}

In the algorithm, we compute for each arc $a$ in $\cal T$ the set of all fingerprints of partial solutions for 
$a$.

\paragraph{Computing sets of fingerprints for leaf arcs}
 We have a separate, simple algorithm for arcs with one endpoint a leaf of $\cal T$.
Let $a = ii'$ be an arc in $\cal T$, with $i$ the parent of leaf $i'$. Note that $V_i$ has at most $2k$ vertices, so $E_a$ has $O(k^2)$ edges. To compute all fingerprints for $ii'$, we can simply enumerate
all $2^{O(k^2)}$ possible orientations of $E_a$, and then check for each if the outdegree weight condition
is fulfilled for all $w\in X_{i'}$, and if so, compute the fingerprint and store it in a table.

\paragraph{Computing sets of fingerprints for other arcs}
Now, suppose $a=ii'$ is an arc with $i$ the parent of $i'$, and $i'$ has at least one child.
Let the children of $i'$ be $j_1, j_2, \ldots, j_q$.
Write $a_p = i'j_p$ for the arc from $i'$ to its $p$th child; $1\leq p\leq q$.

We assume that we already computed (in bottom-up
order) tables $A_{a_p}$ that contain the set of all fingerprints of
partial solutions of $a_p$, for $p\in [1,q]$. 

We now consider the equivalence relation on the arcs $a_1, \ldots a_q$ given by 
$a_{p} \sim a_{p'}$ if and only if $A_{a_p}$ and $A_{a_{p'}}$ are the same, i.e., each fingerprint
that belongs to $A_p$ also belongs to $A_{p'}$ and vice versa.

\begin{claim}
The number of equivalence classes of $\sim$ is bounded by $2^{(k+1)^k}$.
\end{claim}

\begin{claimproof}
Each fingerprint maps at most $k$ vertices to an integer in $[0,k]$, so there are at most
$(k+1)^k$ fingerprints. In a table, each of these can be present or not, which gives the bound. 
\end{claimproof}

(A sharper bound is possible by using that the sum of the values is bounded by $k$.)

We denote the set of all equivalence classes of $\sim$ by $\Gamma$, and denote  
the set of all possible fingerprints for arcs between $i'$ and a child by $\Delta$, i.e.,
$\Delta$ is a subset of the set of all functions $f\colon X_{i'} \rightarrow [0,  k]$.

For an equivalence class $\gamma \in \Gamma$, and fingerprint $f\in \Delta$, we write
$f\in \gamma$ if there exists an arc $a_p$ which belongs to equivalence class $\gamma$ and
$f\in A_{a_p}$. Note that this implies that $f\in A_{a_{p'}}$ for all $a_{p'}$ equivalent to $a_p$.

Let $\rho$ be a partial solution of $a$. The {\em blueprint} of $\rho$ is the function
$g\colon \Gamma \times \Delta \rightarrow [0,q] $, such that
$g(\gamma,f)$ is the number of arcs $a_p$ in equivalence class $\gamma$ for which the restriction of $\rho$ to $E_{a_p}$ 
has fingerprint $f$.

\paragraph{Overall procedure}
The procedure to compute the set of fingerprints for $a$ has a main loop. Here, we enumerate
all orientations $\rho$ of the edges between a vertex in $X_i$ and a vertex in $X_{i'}$ and the edges with
both endpoints in $X_{i'}$. In a subroutine, which will be given later, we check if this orientation
can be extended to a partial solution for $ii'$. If so, we store the fingerprint of this
orientation in the table $A_{ii'}$. If not, the orientation is ignored and we continue with
the next orientation in the ordering.

Note that $\rho$ gives all information to compute the fingerprint, as all edges in $E_a$ that have
an endpoint in $X_i$ receive an orientation in $\rho$.

Assuming the correctness of the subroutine, this gives the complete set of all fingerprints
of $ii'$. Note that the number of edges we orient in this step is bounded by a function of $k$: all these edges are in $(X_i \times X_{i'} \cup X_{i'} \times X_{i'})\cap E$, and thus, its number is
bounded by $2k^2$, and we call the subroutine for at most $2^{2k^2}$ orientations.

\paragraph{The main subroutine}
We now finally come to the heart of the algorithm. In this subroutine, we check whether an orientation
coming from the enumeration as described above can be extended to a partial solution.

To be more precise, we have an arc $a=ii'$ with $i$ the parent of $i'$. Say $i'$ has $q>0$ children,
$j_1, \ldots, j_q$. We are given an orientation $\rho$ of the edges with either one endpoint in $X_i$ and
one endpoint in $X_{i'}$ or with both endpoints in $X_{i'}$. For each child $j_p$, we are given
the table $A_{i'j_p}$ of fingerprints of partial solutions of $i' j_p$.
The procedure returns a Boolean,
that is true if $\rho$ has an extension that is a partial solution of $a= i i'$, and false otherwise.

To do this, we search for the blueprint of such an extension. Let $\Gamma$ and $\rho$ be as above.

First, compute for all equivalence classes $\gamma\in \Gamma$, the number $n(\gamma)$ of arcs to
children of $i'$ that belong to the equivalence class $\gamma$.

Second, we build an integer linear program (ILP). The ILP has a variable $g_{\gamma, f}$ for each
equivalence class $\gamma \in \Gamma$ and  $f\colon X_{i'} \rightarrow [0,k]$ one of the fingerprints stored in tables in class $\gamma$.
This variable denotes the value in the blueprint of the extension that we are searching for.
Furthermore, we have a number of constraints.
\begin{enumerate}
    \item For all $\gamma \in \Gamma$, $f\in \Delta$: $g_{\gamma,f} \geq 0$. (When the variable exists.)
    \item For all $\gamma\in \Gamma$: $\sum_f g_{\gamma,f} = n(\gamma) $. Indeed, the extension has a partial
    solution for each $i' j_p$ whose fingerprint belongs to the equivalence class of $i' j_p$. The
    total number of times such a fingerprint is taken from this equivalence class must be equal to the number of child arcs which belong to the class.
    \item For all $v\in X_{i'}$, we have a condition that checks that the outdegree of $v$ in the orientation belongs to $D_v$. Let $D_v = [d_{min,v},d_{max,v}]$. Let $\alpha$ be the total
    of the weight of all edges in $\rho$ that have $v$ as endpoint and are directed out of $v$.
    Now, add the inequalities:
    \[ d_{min,v} \leq
     \alpha + \sum_{\gamma,f} f(v) \cdot g_{\gamma,f} 
    \leq d_{max,v}
    \]
    where we sum over all $\gamma\in \Gamma$, and $f\in \gamma$.
\end{enumerate}

\begin{claim}
The set of inequalities has an integer solution if and only if there exists a partial solution for
$a$ that extends $\rho$.
\end{claim}

\begin{claimproof}
First, suppose the set of inequalities has an integer solution with values $g_{\gamma,f}$.  Start by assigning to each child arc $a_p$ a fingerprint from $A_{a_p}$ in such a way that
for each equivalence class $\gamma\in \Gamma$, exactly $g_{\gamma,f}$ members of the class have $f$ assigned to it. We can do this because of the second set of inequalities of the ILP.
Then, for each child $j_p$, orient the edges in $A_{a_p}$ as in a partial solution that has $f$ 
as fingerprint --- we can do this, since $f\in A_{a_p}$ by construction. Combine these with the
orientation $\rho$. We claim that this gives a partial solution for $ii'$. For every vertex $v$  belonging
to a bag that is a descendant of $i'$, its outdegree lies in $D_v$, since all its incident edges belong to
a partial solution for an arc between $i'$ and a child. For a vertex $v\in X_i$, the total weight of outgoing edges equals $\alpha + \sum_{\gamma,f} f(v) \cdot g_{\gamma,f}$: $\alpha$ for the edges in
$\rho$, and for each $\gamma$, $f$, there are $g_{\gamma,f}$ children of $i$ which are in equivalence
class $\gamma$ and are oriented with a fingerprint $f$, which implies that the total weight of
outgoing edges from $v$ to vertices in that subtree equals $f(v)$. The third set of conditions of the ILP thus enforces that the total weight of outgoing edges for $v$ is in $D_v$.

In the other direction, suppose we have a partial solution $\rho'$ of $a$ that extends $\rho$. Take
the blueprint of $\rho'$. One can easily verify that setting the variables according to this
blueprint gives a solution of the ILP.
\end{claimproof}

We can now finish the argument.
Note that the number of variables and inequalities is a function of $k$. Thus,
we can solve the ILP with an FPT algorithm, see e.g., the discussion in \cite[Section 6.2]{PA}.
In fact, the number of variables of the ILP is at most
$2^{(k+1)^k} \cdot (k+1)^k$. The number of inequalities is
$O(2^{(k+1)^k})$: we have $O(1)$ inequalities per equivalence class and one for
each vertex in $X_i$. Each integer in the description of the ILP is bounded by
either $k$, or the number of children of $i$. 

Solving ILP's is fixed parameter tractable when we take the number of variables
as parameter, see
\cite[Theorem 6.4]{PA}). We note that the number of variables is double exponential
in $k$; and applying Theorem 6.4 from \cite{PA} gives an algorithm which is quadruple exponential in $k$.

\paragraph{Wrap-up}
All elements of the algorithm have been given. We compute the table $A_a$ bottom-up for each arc $a$ in the partition tree $\cal T$ (e.g., in postorder). A simple procedure suffices when the arc has a leaf as one of its endpoints. Otherwise, we enumerate over orientations of 
edges in $X_{i'}$ and edges between $X_i$ and $X_{i'}$ as described above, 
and for each orientation, we use an ILP to answer the question whether the orientation can be extended to a
partial solution. When the test succeeds, we store the corresponding fingerprint.

When we have finally obtained the table for the arc to the root node, we just check whether this table is non-empty; if so,
we answer positively, otherwise, we answer negatively.

The running time is dominated by the total time of solving all ILP's. 
The number of 
arcs in $\cal T$ is linear in the number of vertices;
for each, we consider $2^{2k^2}$ orientations in the enumeration, and for each of these, the time for
solving the ILP can be done with an algorithm that is fixed parameter tractable in $k$.

We obtain the following result.

\begin{theorem}
{\sc Outdegree Restricted Orientation} is fixed parameter tractable when parameterized by stable gonality, assuming that a finite harmonic morphism of a refinement of the input graph to a tree 
of degree $\sgon(G)$ is given as part of the input; and when parameterized by (stable) treebreadth, assuming a tree partition (of a subdivision) is given realizing the given breadth.
\label{theorem:oroalgorithm}
\end{theorem}

With help of reductions, we obtain fixed parameter tractability for the six other problems parameterized by stable gonality or treebreadth.

\begin{corollary}
The following problems are fixed parameter tractable with stable gonality as parameter, assuming that a finite harmonic morphism of a refinement of the input graph to a tree 
of degree $\sgon(G)$ is given as part of the input; ; and when parameterized by (stable) treebreadth, assuming a tree partition (of a subdivision) is given realizing the given breadth.
\begin{enumerate}
    \item {\sc Target Outdegree Orientation}
    \item {\sc Chosen Maximum Outdegree}
    \item {\sc Minimum Maximum Outdegree}
    \item {\sc Circulating Orientation}
    \item {\sc Undirected Flow with Lower Bounds}
    \item {\sc All-or-Nothing Flow}
\end{enumerate}
\end{corollary}

\begin{proof}
{\sc Target Outdegree Orientation}, {\sc Chosen Maximum Outdegree} and {\sc Minimum Maximum Outdegree}
are special cases of {\sc Outdegree Restricted Orientation}, so we can apply the 
algorithm of Theorem~\ref{theorem:oroalgorithm}. We can also use this algorithm for
{\sc Circulating Orientation}, as this problem is the special case of {\sc Target Outdegree Orientation},
where for each vertex $v$, $d_v$ is set to half the sum of the weights of all edges incident to $v$.

To solve {\sc Undirected Flow with Lower Bounds}, we transform an input of that problem to
an equivalent input of {\sc Circulating Orientation}, as in Lemma~\ref{lemma:uflb2co-sgon}. The breadth 
of the tree partition is still bounded by $4$ times the original value, which yields fixed parameter tractability.

Finally, to solve {\sc All-or-Nothing Flow}, we follow the transformation described by
Lemma~\ref{lemma:aonf2too} --- we obtain an equivalent instance of {\sc Circulating Orientation} with
an associated finite harmonic morphism from a refinement to a tree of bounded degree, and thus
can apply the FPT algorithm for {\sc Circulating Orientation} on that equivalent input.
\qed
\end{proof}

\section{{\sc Capacitated Dominating Set} for graphs with small tree partition width}
\label{section:dominatingset}

In this section, we give the algorithm for {\sc Capacitated Red-Blue Dominating Set}
for graphs with a given tree partition of bounded tree partition width, and discuss at the end of this section how to
adapt this to {\sc Capacitated  Dominating Set}. The description of the algorithm has a large number of technical details.
Before giving these details, we first give a high level overview of the main ideas.

\subsection{A high level overview}
Let $\mathcal{T}$ be the given tree partition of the input graph $G$. The algorithm employs dynamic programming on the tree for $\cal T$. For each
node, we compute a table with the minimum sizes for equivalence classes
of partial solutions. 

To a node $i$, we associate the subgraph $G_i$, consisting of all vertices
in the bags equal to or a descendant of $i$, and the edges in $G$ between these
vertices. A partial solution knows which red vertices in this subgraph belong
to the dominating set, and how blue vertices in this subgraph are dominated.
Possibly the blue vertices in $X_i$ are not yet dominated, as they
can be dominated by vertices from the bag of the parent of $i$.

We have a technical lemma claiming that the differences of the minimum values
for the equivalence classes of a node $i$ is bounded by $2k$. This
lemma is similar to and inspired by a notion called {\em finite integer index} \cite{BodlaenderF01}; the techniques used to prove the lemma are
taken from the classical theory of algorithms to find matchings in graphs (with one
step basically employing the principle of an alternating path). 

If we want to build a partial solution for a node $i$, we decide `what happens around $X_i$', and for each child, we select a partial solution.\footnote{A technicality is that we actually take a slight extension of a partial
solution, but we leave that discussion for Section~\ref{section:Cds-tech}.}

Instead of looking at all children separately, we note that if we take
the table of minimum sizes of a child, and subtract from each the smallest
value, then all remaining values are in the interval $[0,2k]$. 
We sum  these smallest values over all children --- as we need to always
`pay' this amount, we store this value separately.

We now define an equivalence relation on the children of the node --- two
children are equivalent if after subtracting the minima, the table of values 
of minimum sizes are the same. The table size is a function of $k$,
and the values after subtraction are in $[0,2k]$, so this equivalence relation has a number of classes that is bounded by a function of $k$.

Now, instead of deciding for each child separately what type of partial solution we take in its subgraph, we decide for each equivalence class and
each type of partial solution how many children in that equivalence class
have a partial solution of that type. Which of these decisions are possible
and combine to a partial solution, and which gives the minimum total size
for the type of partial solution at $i$ we want to build can be formulated
as an integer linear program (ILP). This ILP has a number of variables
that is a function of $k$, and thus can be solved with an FPT algorithm.

Finally, if we have the table for the root node, it is easy to read off from
it the minimum size of a capacitated red-blue dominating set.

\subsection{Detailed explanation}
\label{section:Cds-tech}

Suppose that we have a graph $G=(V,E)$, with $V=R \cup B$; $R$ is the set of the red vertices, and $B$ the set of blue vertices. We have a capacity function $c\colon R \rightarrow \Z_{>0}$, that gives each red vertex a positive capacity.

\begin{figure}
    \centering
    \includegraphics{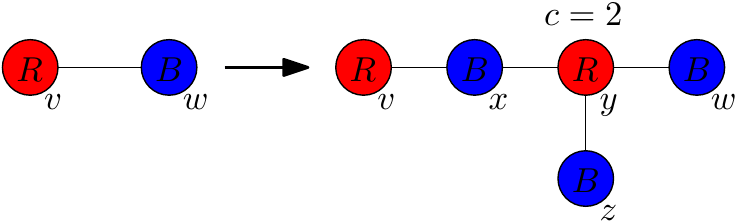}
    \caption{A transformation to handle subdivisions}
    \label{figure:rbds-subdivision}
\end{figure}

\paragraph{Dealing with subdivisions}
If we have a tree partition of a subdivision of $G$, then we can build an equivalent instance where we have a tree partition 
of the graph itself. For each bag that contains
a subdivision vertex,  we replace
this subdivision vertex by three new vertices, as in Figure~\ref{figure:rbds-subdivision}. The three vertices
are placed in the subdivision bag instead of the subdivision vertex. This gives an equivalent
instance:
The new red vertex $y$ has capacity 2, and must be in the dominating set in order to dominate
its private neighbour $z$; if $v$ dominates $w$ in the original graph, then in the transformed graph, $v$ dominates 
$x$ and $y$ dominates $w$; if $v$ does not dominate $w$ in the original graph, then another vertex dominates $w$, and
$y$ can be used to dominate $z$ and $x$. The width of the 
tree partition at most triples.

\medskip

From now on, we assume that we are given a tree partition $\cal T$ of $G=(R\cup B,E)$
of width $k$, and a weight function $w: R \rightarrow \Z_{>0}$.

View the tree of $\cal T$ as a rooted tree. For a bag $X_i$, let $V_i$ be the set of vertices
in the bags $X_j$ with $j=i$ or $j$ a descendant of $i$, $E_i$ be the set of edges in
$E$ with both endpoints in $V_i$, and $G_i = (V_i,E_i)$.

\paragraph{Solutions}
A {\em solution} is a pair $(S,f)$, with $S\subseteq R$ a set of
red vertices in $G$, and $f\colon B \rightarrow S$ a mapping that assigns each
blue vertex to a neighbour, such that the capacity constraints are
satisfied: for all $v\in S$, $|f^{-1}(v)|\leq c(v)$. 

One can check in polynomial time whether at least one solution exists: if there
is a solution, then we can take $S=R$, so what is needed is to see whether 
we can assign all blue vertices to red neighbours satisfying the capacities; this step can be done by a standard generalized matching algorithm.

\paragraph{Partial solutions}
We will use three different, but slightly similar, notions that describe a part
of a solution restricted to a subgraph: partial solutions, 
partially extended partial solutions (peps), and extended partial solutions (eps). While it may seem somewhat confusing
to have three such notions, this
helps for a clear explanation of the algorithm and the proof of its correctness.

For each of these three notions, we define an equivalence relation by identification of a `characteristic', 
and we define the minimum `size' of an element in an equivalence class (for convenience, assigning size $\infty$ when no solution with the given characteristic exists).

We start with giving these definitions for partial solutions.

Suppose $i$ is a node from $\cal T$.

A \textit{partial solution} for $i$ is a triple $(S,D,f)$, where $S\subseteq R\cap V_i$ is a set of red vertices in $V_i$, $D \subseteq B \cap X_i$ is a set of blue vertices
from $X_i$, and $f\colon B \cap (V_i\setminus X_i) \cup D \rightarrow S$ is a function, that maps each
blue vertex in $V_i \setminus X_i$ or in $D$ to a vertex in $S$, such that for all $v \in B \cap (V_i\setminus X_i) \cup D $, $f(v)$ is a neighbour of $v$, and for all $w\in S$, $|f^{-1}(w)|\leq c(w)$.

In other words, in the partial solution, we need to dominate all blue vertices in sets $X_{i'}$ 
with $i'$  a descendant of $i$, and 
from $X_i$, we need to dominate all vertices in $D$ but not those in $X_i \setminus D$.
This domination is done by the vertices
in $S$ without exceeding the capacities.

Note that a partial solution $(S, X_r\cap B, f)$ for root bag $r$ of $\cal T$ is
a solution to the \textsc{Capacitated Red-Blue Dominating Set} problem, and we thus
want to determine the minimum size of a set $S$ such that there is a partial solution
$(S,X_r \cap B, f)$ for the root bag. 

The \emph{characteristic} of a partial solution $(S,D,f)$
is the pair $(D,h)$, with $h\colon R\cap X_i \rightarrow [0,k]$ the function
such that for $v\in S$ we have 
$h(v) = \min \{ k, c_f(v)\}$ with $c_f(v) = c(f) - |f^{-1}(v)|$,
and for $v\not\in S$, we set $h(v)=0$.
The value $c_f(v)$ is what remains from the capacity of $v$ after all vertices
are mapped to $v$ by $f$. When we extend a partial solution,
the only additional vertices that are mapped to vertices in $X_i$ are
those from the parent bag of $i$, and thus, we will not need more than
$k$ additional capacity --- hence, we can take the maximum of the remaining
capacity and $k$. Vertices not in $S$ cannot have additional vertices
mapped to it, so their remaining capacity is set to $0$.

We say that two partial solutions for $i$ are equivalent if and only
if they have the same characteristic.

The {\em size} of a partial solution $(S,D,f)$ is $|S|$. The {\em minimum
size} of an equivalence class is the minimum size of a partial solution
in the class; we denote by $A_i(D,h)$ the minimum size of the equivalence
class of partial solutions at $i$ with characteristic $(D,h)$.

We say that a solution $(S,f)$ \emph{extends} a partial solution $(S',D,f')$
when $S'= S\cap V_i$ and 
$f'$ is obtained from $f$ by restricting the domain to
$((V_i \setminus X_i)\cap B) \cup D$, with the additional properties that blue
vertices in $X_i \setminus D$ are mapped to the parent bag:
if $v\in (X_i \setminus D)\cap B$, then $f(v)\not\in V_i$.

Exchange arguments show that a partial solution that is not of minimum
size in its equivalence will never extend to a solution of minimum size.
Thus, in the dynamic programming algorithm, we need to compute
for all the equivalence classes of partial solutions their minimum
size.

\paragraph{Partial extended partial solutions}
For nodes that are not the root of $\cal T$, we define the notion
of partial extended partial solution. We use these to prove
a property of values of minimum sizes which is needed in the dynamic
programming algorithm.

Suppose $i$ is not the root of $\cal T$, and let $j$ be the parent of $i$.
A \textit{partial extended partial solution} or \emph{peps} for $i$ is a triple
$(S,D,f)$, where $S\subseteq R\cap V_i$ is a set of red vertices in $V_i$,
$D \subseteq B \cap (X_i \cup X_j)$, and 
$f\colon B \cap (V_i\setminus X_i) \cup D \rightarrow S$ is a function that maps each
blue vertex in $V_i \setminus X_i$ or in $D$ to a vertex in $S$, such that for all $v \in B \cap (V_i\setminus X_i) \cup D $, $f(v)$ is a neighbour of $v$, and for all $w\in S$, $|f^{-1}(w)|\leq c(w)$.

The only difference between partial solutions and partial extended partial solutions
is that a peps knows which blue vertices in $X_j$ are dominated
by red vertices in $X_i$.

The {\em characteristic} of a peps $(S,D,f)$ is $D$. 
Two peps are equivalent when they have the same characteristic.
The \emph{size} of a peps $(S,D,f)$ is $|S|$, and the \emph{minimum size} of
an equivalence class of peps is the minimum size of a peps in the class.
We denote this by $B_i(D)$ for the class with characteristic $D$.

A solution $(S,f)$ extends a peps $(S',D,f')$, when 
$S'= S\cap V_i$ and $f'$ is the restriction
of $f$ to 
 $((V_i \setminus X_i)\cap B)\cup D$ 
and every blue vertex in $X_i \setminus D$ is mapped to a vertex in $X_j$ and every blue vertex in $X_j\setminus D$ is not mapped to a vertex in $X_i$.
(I.e., the peps tells for all red vertices in $V_i$ which
blue vertices they dominate.)
Again, a peps that is not of minimum size in its equivalence class
cannot be extended to a minimum size solution, and thus, the dynamic
programming algorithm needs only store the minimum size for each equivalence class of peps.

\paragraph{Extended partial solutions}
Again, let $i$ be a node that is not the root, with parent $j$.

An \textit{extended partial solution} or \textit{eps} for $i$ is a triple
$(S,D,f)$, where $S\subseteq R\cap V_j$ is a set of red vertices in $V_j$,
$D \subseteq B \cap X_j$, and 
$f\colon (B \cap V_i) \cup D \rightarrow S$ is a function that maps each
blue vertex in $V_i$ or in $D$ to a vertex in $S$, such that
for all $v\in (B \cap V_i) \cup D$, $f(v)$ is a neighbour of $v$, 
and for all $w\in S$, $|f^{-1}(w)|\leq c(w)$, and for all $x\in D$, $f(x)\in V_i$.

The difference between a partial extended partial solution and an extended partial
solution is that in the latter, all blue vertices in $X_i$ are dominated, possibly
by red vertices from $X_j$. Blue vertices in $X_j$ can but do not have to be
dominated, and we only look at domination of these vertices by red vertices from $X_i$. In the characteristic, we record how much capacity of the
red vertices in $X_j$ is used to dominate the blue vertices in $X_i$.

Note that at this point, the eps only tells us which blue vertices in $X_i$ are
dominated by red vertices in $X_j$; in an extension, these red vertices
can dominate other vertices not in $V_i$.

The \emph{characteristic} of an eps $(S,D,f)$ is 
the pair $(D,g)$ with $g: X_j \cap R \rightarrow [0,k]$ is given
by $g(v) = | f^{-1}(v)|$. Two eps are equivalent if they have the same
characteristic.

In the size of an eps, we do not yet count the red vertices in $X_j\cap S$
--- this helps to prevent to count these multiple times, and makes later steps
slightly simpler.
The \emph{size} of an eps $(S,D,f)$ at $i$ is $|S \setminus X_j|$.
The \emph{minimum size} of an equivalence class of eps is the minimum size
of an eps in the class. We denote the minimum size of an eps at $i$ with
characteristic $(D,g)$ by $C_i(D,g)$.

A solution $(S,f)$ is an extension of an eps $(S',D,f')$ when $S'\cap V_i = S \cap V_i$? and
$f'$ is obtained by restricting the domain of $f$ to the union
of the blue vertices in $V_i$ and the blue vertices in $X_j$ that are
dominated by vertices in $X_i$. Again, an eps that is not of minimum
size in its equivalence class has no extension of minimum size, and thus,
in the dynamic programming algorithm, we need to tabulate only the minimum
sizes of equivalence classes of eps.

\paragraph{The finite integer index property}
Before we proceed with the algorithm, we need a technical lemma, Lemma~\ref{lemma:fii}. This lemma is used
to show that a certain equivalence relation has a finite number of equivalence classes.
This shows that the problem is  finite integer index in the terminology of ~\cite{BodlaenderF01}; we do not use the terminology from that source any further, but we note that our methods are similar to the ones in that reference. 
The proof of the lemma is heavily inspired by well known techniques from matching, in
particular, the notion of an alternating path, and the proof that such exist.

\begin{lemma}
Let $i$ be a node.
If there is no peps $(S, \emptyset, f)$ for any $S$ and $f$ for $i$, then
no solution exists.
\end{lemma}

\begin{proof}
Suppose we have a solution for $G$, say with dominating set $S'$, and $f'$
assigns each blue vertex to a neighbour in $S$ without violating capacities.
Then set $S=S'\cap V_i$, and $f$ the restriction of $f'$ to domain $(V_i \setminus X_i)\cap B$.
\end{proof}

From now on, we assume that a peps 
$(S, \emptyset, f)$ for some $S$ and $f$ for $i$ exists.

\begin{lemma}\label{lemma:fii}
Let $i$ be a node with parent $j$.
Suppose $\alpha = B_i(\emptyset)$. Let $D\subseteq (X_i\cup X_j)\cap B.$
If there is a peps with characteristic $D$, 
then $$\alpha \leq B_i(D) \leq \alpha + |D|.$$
\end{lemma}

\begin{proof}
We have that $\alpha \leq B_i(D)$, as we can take a solution for $D$, and
restrict the domination function $f$ by removing $D$ from the domain.

We prove that $B_i(D) \leq \alpha + |D|$
by induction with respect to the size $|D|$ of $D$. The result trivially holds when $|D|=0$.

Consider a $D\subseteq X_i\cup X_j$ with $|D|>0$, and suppose the lemma holds for
smaller sized sets. Suppose there exists a peps $(S,D,f)$ for some $S$ and $f$; if not,
we are done. Take a vertex $v\in D$. Let $f\setminus v$ be the restriction of
$f$ where we remove $v$ from the domain of $f$. Now, $(S,D\setminus \{v\}, f\setminus v)$
is a peps.

Thus, the minimum size of a set $S'$ for which there is a peps of the form
$(S', D\setminus \{v\}, f')$ exists for some $f'$, and from the induction hypothesis,
it follows that $|S'|\leq |D\setminus \{v\}|= |D|-1$.

We now define an auxiliary graph $G^+$, which is formed from $G$ by replacing each
red vertex $v$ by $c(v)$ copies of $v$, each incident to all neighbours of $v$.
To a peps $(S'', D'', f'')$, we associate a matching in $G^+$ as follows: 
for each vertex $x$ in the domain of $f''$, we add to the matching
an edge from $x$ to a copy
of $f(x)$. As the size of a preimage of a red vertex is at most its capacity,
we can add these edges such that they form a matching, i.e., have no common endpoints.

Let $M_1$ be the matching associated with the peps $(S,D,f)$ and 
let $M_2$ be the matching associated with the peps $(S', D\setminus \{v\}, f')$.
Consider the symmetric difference $M_1 \oplus M_2 = (M_1 \cup M_2) \setminus (M_1 \cap M_2)$.

Vertices in the symmetric difference of two matchings have degree at most two, so
this symmetric difference consists of a number of cycles and paths. 
The vertex $v$ is incident to an edge in $M_1$ but not in $M_2$, so is an endpoint of a path in 
$M_1 \oplus M_2$. All other blue vertices are incident to 0 or to 2 edges in
$M_1\cup M_2$, so have degree 0 or 2 in the symmetric difference, hence cannot be
endpoint of a path. We find that in $M_1 \oplus M_2$ there is a path from $v$ to a red vertex.
Say this last red vertex is $z$, and let $M'_1$ and $M'_2$ be the edges
from $M_1$ and $M_2$ that belong to this path.
We thus have a path $P$ that starts at $v$, and then alternatingly has an edge in $M_1$
from a blue vertex to a red vertex, and an edge in $M_2$ from a red vertex to a blue
vertex, ending with an edge in $M_1$.

We now change the peps $(S', D\setminus \{v\}, f')$ as follows. 
Set $S'' = S' \cup \{z\}$. The domain of $f''$ is obtained by adding $v$ to the domain of $f'$.
All blue vertices that are not on $P$ are mapped in the same way by $f''$ as by $f'$.
For each edge in $M'_1$, say from blue vertex $x$ to a copy of red vertex $y$,
we map $x$ to $y$. This effectively cancels the mappings modelled by the edges in $M'_2$.

We claim that $(S'\cup \{z\}, D,f'')$ is a peps.\footnote{This step is very similar to
the use of an alternating path to augment a flow, as in classical network flow theory.}
Indeed, consider a  red vertex on $P$, unequal to $z$; say it is a copy of the red vertex $y$.
The vertex has an incident edge in $M'_1$ (which causes one
additional mapping to this vertex) and an incident edge in $M'_2$ (which cancels one mapping to
this vertex), and thus, the total number of blue vertices dominated by $y$ stays the same, in particular, 
at most $c(y)$. Since $z$ has no incident edge in $M'_2$, it is a copy of a vertex that has at least one
capacity left in $f'$.

As $(S'\cup \{z\}, D,f'')$ is a peps, $B_i(D) \leq B_i(D-\{v\})+1 \leq \alpha + |D|$,
and the induction step
is proved.
\end{proof}

\begin{lemma}\label{lemma:fii2}
Let $(D_1,g_1)$ and $(D_2,g_2)$ be characteristics of an eps at $i$.
Then $$|C_i(D_1,g_1) - C_i(D_2,g_2)| \leq 2k.$$

\end{lemma}

\begin{proof}
Each eps of minimum size at $i$ can be obtained from a peps of minimum size
at $i$ by extending it by dominating the not yet dominated vertices
from $X_i$ by red vertices in $X_j$. Note that we do not count the red
vertices in $X_j$ in the sizes of eps at $i$. Thus, the largest difference
in sizes of eps at $i$ is bounded by the largest difference in sizes of
peps at $i$, which, by Lemma~\ref{lemma:fii}, is at most $2k$.
\end{proof}

Suppose the problem has at least one solution for $G$.
By Lemma~\ref{lemma:fii}, for each node $i$ and $D\subseteq (X_i \cup X_j)\cap B$, we have $B_i(D) \in [B_i(\emptyset), B_i(\emptyset)+2k]$.

\paragraph{Main shape of algorithm}
We now describe the algorithm.

We start with a generalized matching algorithm to determine whether the
set of all red vertices is a capacitated dominating set. If not, 
there is no solution, and we halt. Otherwise, all values $B_i(\emptyset)$ 
are integers (and at most $|R|$).

We then process all nodes in postorder. Each computation step below is done by a subroutine whose details are discussed further below. 
For a leaf node, we do a direct computation of the table with
the minimum sizes of equivalence classes of partial solutions ($A_i$).
For a node with children, we use the tables with minimum sizes of
equivalence classes of
eps of its children ($C_{j_\alpha}$ for all children $j_\alpha$) to
compute the table with minimum sizes of partial solutions ($A_i$).
If the node is the root, we decide the problem.
If the node is not the root, we next compute a table with minimum sizes
of equivalence classes of peps ($B_i$) and then a table
with minimum sizes of equivalence classes of partial solutions ($C_i$).
We are now done processing this node, and continue with the next.

\paragraph{Partial solutions of leaf nodes}
A simple exhaustive search gives all partial solutions for a leaf node,
since when $i$ is a leaf of $\cal T$, then $G_i$ has at most $k$ vertices, and
$O(k^2)$ edges.

\paragraph{Partial solutions for nodes with children}
Suppose now $i$ is a node in $\cal T$ with at least one child.
Let $j_1$, \ldots, $j_q$ be the children of $i$. We describe how
to obtain a table with minimum sizes for equivalence classes of partial
solution for $i$, given tables with minimum sizes for equivalence classes of 
extended partial solutions for the children of $i$. This is the most
involved (and most time consuming) step of the algorithm;
the step uses as a further subroutine 
solving an integer linear program with the number of variables a
function of $k$.

We assume we have computed, for each child $j_\alpha$ and each equivalence class of extended partial solutions for $j_\alpha$, the minimum
size of the set $|S|$ in such an eps; i.e.,
for all equivalence classes, represented by $(D,g)$, we know the 
value $C_{j_\alpha}(D,g)$.

For each child $j_\alpha$, let $m_\alpha = \min_{D,g} C_{j_\alpha}(D,g)$,
and let $C'_{j_\alpha} (D,g) = C_{j_\alpha}(D,g) - m_\alpha$, if
$C_{j_\alpha}(D,g) \neq \infty$; if $C_{j_\alpha}(D,g) = \infty$,
then we set $C'_{j_\alpha} (D,g) = \infty$. 

By Lemma~\ref{lemma:fii2}, we have $C'_{j_\alpha} (D,g) \in [0,2k] \cup \{\infty\}$, for all children $j_\alpha$ and all characteristics
of eps at $j_\alpha$ $(D,g)$.

We now define another equivalence relation, this time on the children
of $i$, and say that $j_\alpha \sim j_\beta$ if for all $(D,g)$, we have 
$C'_{j_\alpha} (D,g) = C'_{j_\beta} (D,g)$.
 
The index of this equivalence relation is bounded by a (double exponential)
function of $k$.
The domain size of functions $C'_{j_\alpha}$ is bounded by
$2^k \cdot (k+1)^k$, and the image size is bounded by $2k+2$.

Each equivalence class is identified by
a function $C'$ that maps pairs of the form $(D,g)$ to a value
in $[0,2k]\cup \{\infty\}$.

The first step is to compute the number of children in each equivalence class of the
equivalence relation $\sim$ described above. If $C'$ is a function mapping pairs $(D,g)$ to values
in $[0,2k]\cup \{\infty\}$, we write $N[C']$ for the number of children
$j_\alpha$ with $C'_{j_\alpha}= C'$.

We also compute the integer $m_{\mathrm{tot}} = \sum_{\alpha=1}^q m_\alpha$.

Note that for a child $j_\alpha$, the function $C_{j_\alpha}$
captures all essential information pertaining to the subgraph $G_{j_\alpha}$.
Thus, the function $N$ together with the integer $m_{\mathrm{tot}}$
gives at this stage all essential information
for all children of $i$. At this point, we can forget further information
and tables for the children of $i$, and keep only the function $N$
and the value $m_{\mathrm{tot}}$ in
memory.

We now iterate over all characteristics of partial solutions $(D,h)$ at $i$,
and compute the minimum size of the characteristic, as described below.

We then assume $(D,h)$ to be fixed and iterate over all guesses which
red vertices in $X_i$ belong to $S$; say this set is $Q$. (Each iteration yields the
minimum size for the characteristic, which is a non-negative integer, or $\infty$. We keep the best value over all iterations.)

For simplicity, in the remainder of this iteration, we assume that 
vertices in $(X_i\cap R)\setminus Q$ (i.e., red vertices that are not used in
the dominating set) have capacity $0$.

The (blue) vertices in $D$ can be dominated by vertices in $X_i$ or in
a child of $X_i$. We iterate over all possible options.
If we guess that $v\in D$ is dominated by a vertex $w\in X_i$, then
subtract one from the capacity of $w$ (and reject this case when it leads to a negative capacity), and remove $v$ from $D$.
All vertices that remain in $D$ must be mapped to vertices in child bags.

\paragraph{An ILP for characteristic selection}

We now want to determine if we can choose from each child bag a characteristic
of an eps such that the combination of these characteristics and the
guesses done so far combine to a partial solution with characteristic
$(D,h)$; and if so, what is its minimum size.

We can do this in FPT time by formulating the question as a linear program,
with the number of variables a function of $k$.

For each equivalence class of $\sim$ and each eps at children of $i$,
we have a variable. The variable $x_{C',(D,g)}$ denotes the number
of children in the equivalence class with function $C'$ for which the restriction of the partial solution to an eps for that child
has characteristic $(D,g)$. 

We have a number of constraints.
\begin{enumerate}
    \item For each $x_{C',(D,g)}$, $0 \leq x_{C',(D,g)}$.
    \item  For each $x_{C',(D,g)}$, if $C'(D,g)=\infty$, then
    $x_{C',(D,g)} =0$. In this case, we have subtrees where $(D,g)$ is not
    the characteristic of an eps, so we cannot choose such eps in those subtrees.
    \item For each $C'$, $\sum_{(D,g)} x_{C',(D,g)} = N[C']$. This enforces
    that we choose for each child in the equivalence class for $C'$ precisely one extended partial solution.
    \item For each $v\in D$, $\sum_{C',(D,g), v\in D} x_{C',(D,g)} =1 $. This
    enforces that each vertex in $D$ is dominated exactly once by a vertex from a child node.
    \item For each $v\in X_i\cap R$ with $h(v) < k$:
    $c(v) - \sum_{C',(D,g)} g(v) \cdot x_{C',(D,g)}
    = h(v)$. The sum tells how much capacity of $v$ is used by vertices in
    child bags of $v$; $h(v)$ is the exact remaining capacity. This case and the next enforce that $h$ is the minimum of the remaining capacity and $k$.
    \item For each $v\in X_i\cap R$ with $h(v)= k$:
    $c(v) - \sum_{C',(D,g)} g(v) \cdot x_{C',(D,g)}
    \geq k$. This case is similar as the previous one. We have a different
    case here because we took the minimum of $k$ and the remaining capacity
    for functions $h$ in the characteristics of partial solutions.
\end{enumerate}

The total size of the partial solutions equals 
$$ |Q| + m_{\mathrm{tot}}+ \sum_{C',(D,g)} C'(D,g) \cdot x_{C',(D,g)}. $$
Indeed, we have $x_{C',(D,g)}$ many children in the class of $C'$ 
where we choose an eps
with characteristic $(D,g)$. Consider one of these, say $j_\alpha$.
The number of vertices in the dominating set in $V_{j_\alpha}$ is
the sum of $m_{j_\alpha}$ and $C'(D,g)$. We have $m_{j_\alpha}$
counted within $m_{\mathrm{tot}}$, and $C'(D,g)$ is counted in the summation,
once for each time we choose this eps for this equivalence class.

Thus, in the ILP, we minimize $\sum_{C',(D,g)} C'(D,g) \cdot x_{C',(D,g)}$,
and report for this iteration, if there is at least one solution of the ILP,
 the sum of the minimum value of the ILP
and $m_{\mathrm{tot}}+|Q|$.

\paragraph{From partial solutions to peps}
Suppose we have a table of values $A_i(D,h)$ for node $i$ with parent $j$.
To compute the table of values $B_i(D)$, we do the following.
For each set $D\subseteq (X_i\cup X_j) \cap B$ of blue vertices in $X_i\cup X_j$
(i.e., characteristics of peps at $i$)
consider all characteristics of partial solutions at $i$ of the form
$(D\cap X_i, h)$. Check all mappings of the vertices in $D\cap X_j$ to
red neighbours in $X_i$. For each such mapping, we check whether we obtain
a peps (necessarily with characteristic $D$) if we extend a partial
solution with characteristic $(D\cap X_i, h)$ using this mapping; and if
so, we check whether the size $A_i(D,h)$ is the current best for the characteristic $D$. (The check amounts to verifying that red
vertices $v\in X_i\cap R$ have at most $h(v)$ vertices from $X_j$ mapped to it. Note that red vertices in $X_i$ that are not in $S$ have remaining
capacity $h(v)=0$ so will not dominate blue vertices in $X_j$.)

\paragraph{From peps to eps}
Suppose we have a table of values $B_i(D)$ for node $i$ with parent $j$.
To compute the table of values $C_i(D,g)$, we do the following.

For each characteristic $(D,g)$ of an eps at $i$, look at all sets
$D'\subseteq X_i\cap B$ of blue vertices in $X_i$. Check if 
there exists a mapping of 
the vertices in $(X_i\cap B) \setminus D'$ to red neighbours in $X_j$,
such that for each vertex $v\in X_j\cap R$ exactly $g(v)$ vertices are mapped
to it. If this is so, we can extend a peps with characteristic $D'$ to
an eps with characteristic $(D,g)$; the size stays the same, since in an eps,
vertices in $X_j$ do not yet contribute to the size. The minimum size
for the characteristic $(D,g)$ is the minimum over all the $B_i(D')$ for
all $D'$ which have such a mapping.

\paragraph{Finding the answer}
A partial solution $(S,D,f)$ at the root node $r$ gives a solution
$(S,f)$ if $D=X_r\cap B$ --- in that case, all vertices, including
all blue vertices in the root bag are dominated. Thus,
the minimum size of a solution $(S,f)$ is the minimum size over
all equivalence classes of partial solutions at $r$ with
a characteristic of the form $(X_r \cap B, f)$. We thus answer the problem by computing $\min_f A_r(X_r\cap B, f)$, after we computed the
table $A_r$ for the root $r$.

\paragraph{Time analysis}
We note that for each of the equivalence relations (on partial solutions,
peps, and eps), the index is a function of $k$. Also, the equivalence
relation on children of a node is of finite index, and thus, we solve
an ILP with the number of variables a function of $k$.
Again, we use that solving ILP's with the number of variables as
parameter is fixed parameter tractable; see e.g., \cite[Theorem 6.4]{PA}.
This all amounts
to an FPT algorithm. 

We can now summarize our findings in the following result.

\begin{theorem}
{\sc Capacitated Red-Blue Dominating Set} is fixed parameter tractable when parameterized by tree partition width, assuming that a tree partition realizing the width is given as part of the input (in particular, it is also FPT when parameterized by (stable) tree breadth and stable gonality, assuming a corresponding refinement of the graph with a tree partition or morphisms to a tree is gievn as part of the input). 
\label{theorem:crbdsoalgorithm}
\end{theorem}

\subsection{Capacitated Dominating Set}
It is not hard to deduce the same result for {\sc Capacitated Dominating Set}, either by 
adapting the previous algorithm, or --- somewhat easier --- by using a transformation of the
instance as follows.

\begin{lemma}
Suppose we have an instance $(G,c)$ of \textsc{Capacitated Dominating Set} and a tree partition of a refinement of $G$ of breadth $k$. Then one can build in polynomial time an equivalent instance of \textsc{Capacitated Red-Blue Dominating Set} $(G',c')$ and a tree partition of a refinement of $G'$ of breadth $2k$.
\end{lemma}

\begin{proof}
Replace each vertex $v$ by a red vertex of capacity $c(v) + 1$ and a blue vertex; the red vertex
is incident to its blue copy and all blue copies of neighbours of $v$. The breadth thus precisely doubles.
\end{proof}

\begin{corollary}
{\sc Capacitated Dominating Set} is fixed parameter tractable when parameterized by tree partition width, assuming that a tree partition realizing the width is given as part of the input (in particular, it is also FPT when parameterized by (stable) tree breadth and stable gonality, assuming a corresponding refinement of the graph with a tree partition or morphisms to a tree is gievn as part of the input). 
\label{corollary:cdsoalgorithm}
\end{corollary}

\section{Hardness results}
\label{sec:hardness}

\subsection{XNLP-completeness for bounded pathwidth}

In this section, we  first consider the {\sc All-Or-Nothing Flow} problem with the pathwidth as parameter, and
show that this problem is complete for the class XNLP. After that, we use the transformations given
in Section~\ref{section:transformations} to obtain XNLP-hardness for the the six other problems in Figure \ref{fig:transformations}(b).

XNLP is the class of parameterized problems that can be solved with a non-deterministic algorithm
in $O(f(k)n^c)$ time and $O(f(k)\log n)$ space, with $f$ a computable function, $c$ a constant,
and $n$ the input size. This class was studied in 2015 by Elberfeld et al.~\cite{ElberfeldST15} under the name
$N[f \text{\,poly}, f \log]$. Recently, Bodlaender et al.~\cite{BodlaenderGNS21} showed that a large number
of problems is complete for this class. 

We now describe one of these problems, that we subsequently use to show hardness of {\sc All-Or-Nothing Flow}. 

A {\sc Non-deterministic Non-decreasing Checking Counter Machine} (NNCCM) is a machine model where we have $k$ counters, an upper bound $B \in \Z_{\geq0}$, and a sequence of $n$ tests. Each test
is described by a 4-tuple of integers in $[1,k] \times [0,B] \times [1,k] \times [0,B]$.
The machine works as follows. Initially, all counters are $0$. We  can increase some (or all, or none) of the counters, to integers that are at most $B$ by alternating the following steps:
\begin{itemize}
    \item Each counter is possibly increased, i.e., for each $i$, if the current value of counter $i$ equals $c_i$, then the counter is non-deterministically given a value in $[c_i,B]$.
    \item The next check is executed. If the next check is the 4-tuple $(i,a,j,b)$, then the machine halts and rejects if counter $i$ has value $a$ and counter $j$ has value $b$.
\end{itemize}
If the machine has completed all tests without rejecting, then the machine accepts.

For a test $(i,a,j,b)$, we say that a counter $i'$ {\em participates} in the test if $i'\in \{i,j\}$, 
and we say it {\em fires}, if $i=i'$ and the value of counter $i$ equals $a$, or $j=i'$ and the
value of counter $j$ equals $b$. We consider the following problem.

\begin{verse}
{\sc Accepting NNCCM}\\
{\bf Given:} A Non-deterministic Checking Counter Machine, with $k$ counters, upper bound $B$, and a sequence
of $n$ tests in $[1,k] \times [0,B] \times [1,k] \times [0,B]$, with all values given in unary.\\
{\bf Question:} Is there an accepting run of this machine? 
\end{verse}

\begin{theorem}
{\sc All-or-Nothing Flow} for graphs given with a path decomposition, with the width of the given path decomposition as parameter, is complete for \textup{XNLP}.
\label{theorem:allornothingxnlp}
\end{theorem}

\begin{proof}
Membership in XNLP is easy to see: go through the path decomposition from left to right; for each edge between 
two vertices in the current bag with at least one endpoint not in the previous bag, non-deterministically
choose the flow over the edge. For all vertices in the bag, keep track of the difference between the inflow
and outflow. For a vertex $v$ in the current bag that is not in the next bag, check whether the difference between
the inflow and outflow is 0 if $v\not\in \{s,t\}$, $-R$ if $v=s$, and $R$ if $v=t$, and if not, reject. Accept when all bags are handled.

We have in memory the current value of the $k$ counters, and a pointer to the current
check to be performed; these all can be written with $O(\log n)$ bits.

We next show hardness, with a reduction from {\sc Accepting NNCCM}, which, we recall, was shown to be XNLP-complete in \cite{BodlaenderGNS21}.

Suppose we are given the NNCCM, with
$k$ counters, an upper bound $B$, and a sequence $S$ of $n$ checks, with the
$t$-th check whether counter $c_{t,1}\in [1,k]$ equals $v_{t,1}\in [0,B]$ and counter $c_{t,2}$ equals $v_{t,2}$.

We construct the following all-or-nothing flow network $G=(V,E)$ with capacity function $c$, see also Figure \ref{fig:aonflow-xnlp}.

For each counter $j\in [1,k]$ and each time $t\in [0,n]$, we take a vertex $v_{j,t}$ and a vertex $w_{j,t}$.
We have two additional vertices $s$, $t$, and for each $i$-th check, we have two vertices $x_{i,1}$, $x_{i,2}$. 

\begin{figure}[t]
    \centering
    \includegraphics[width=\textwidth]{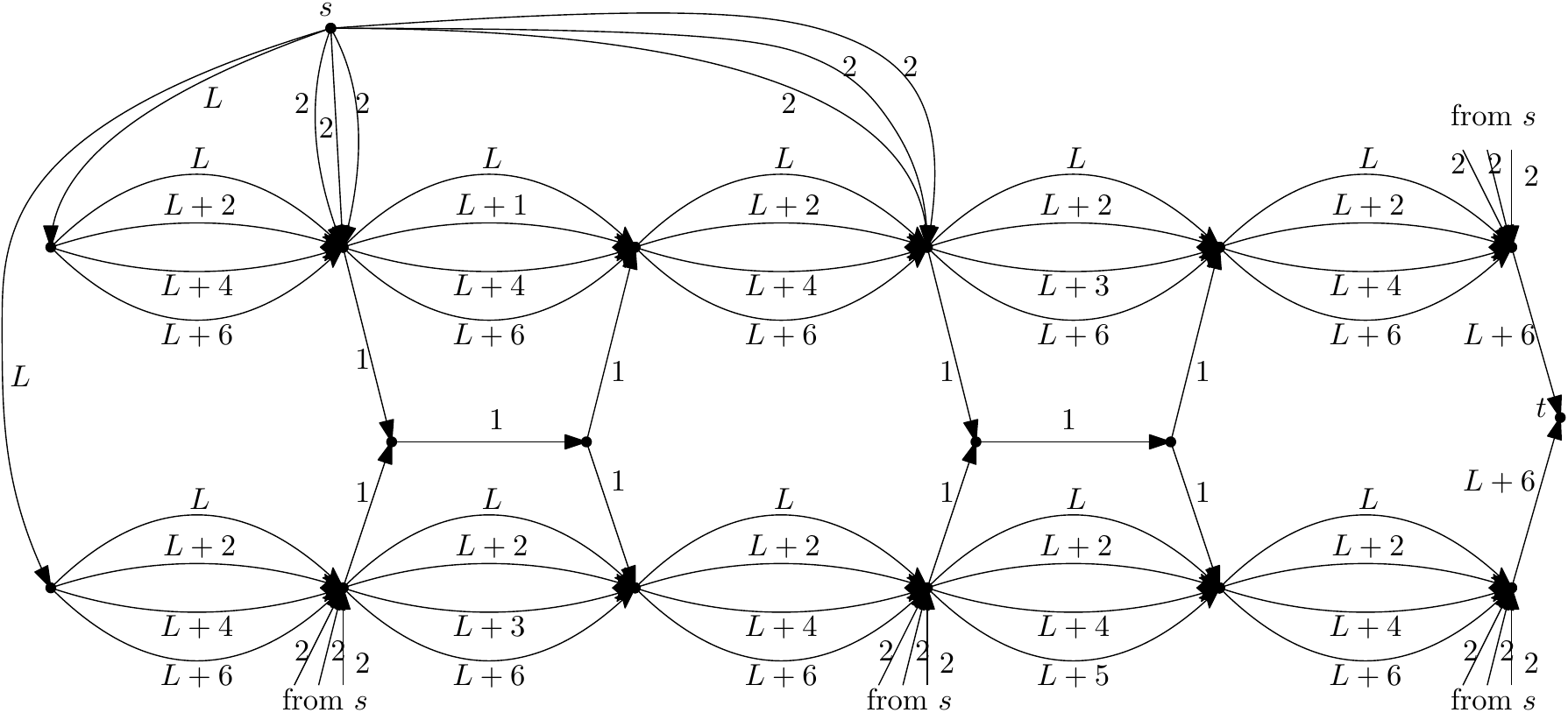}
    \caption{An example of the construction, with 2 counters, $B=3$, and two tests: counter 1 is 1 and counter 2 is 2; and counter 1 is 2 and counter 2 is 3.}
    \label{fig:aonflow-xnlp}
\end{figure}

In the description below, we allow for parallel arcs. We can easily remove these by subdividing each parallel arc.
Notice that such subdivisions increase the pathwidth of the graph by at most one in total.

We set $L = 4knB$ and let $R= k\cdot (L+2B)$. We construct the following arcs with capacities:
\begin{itemize}
    \item For each $j \in [1,k]$, we have an arc from $s$ to $v_{j,0}$ with capacity $L$.
    \item For each $j\in [1,k]$, we have an arc from $w_{j,n}$ to $t$ with capacity $L+ 2B$.
    \item For each $j\in [1,k]$, $t\in [0,n]$, we have $B$ parallel arcs with capacity $2$ from
    $s$ to $w_{j,t}$.
    \item For each $j\in [1,k]$, $t\in [0,n]$, we have a $B+1$ parallel arcs with different capacities
    from $v_{j,t}$ to $w_{j,t}$, with for each $\alpha\in [0,B]$ one arc with capacity $L+2\alpha$.
    \item For each $j\in [1,k]$, $t\in [0,n-1]$, we have a $B+1$ parallel arcs with different capacities
    from $w_{j,t}$ to $v_{j,t+1}$, namely, for each $\alpha \in [0,B]$, we have one arc with weight
    $L+2\alpha$ or $L+2\alpha-1$. The weight of the arc equals $L+2\alpha-1$ if and only if counter $j$ participates in the $(t+1)$st check, and checks if counter $j$ equals $\alpha$. In all other cases, the
    weight of the arc equals $L+2\alpha$. 
    \item For each $i\in [1,n]$, recall that the counters involved in the $i$-th check are
    $c_{i,1}$ and $c_{i,2}$. Take arcs with capacity 1 from $w_{c_{i,1},i-1}$ to $x_{i,1}$,
    $w_{c_{i,2},i-1}$ to $x_{i,1}$, from $x_{i,1}$ to $x_{i,2}$,
    from $x_{i,2}$ to $v_{c_{i,1},i}$, and from $x_{i,2}$ to $v_{c_{i,2},i}$.
\end{itemize}

\begin{lemma}
There is an all-or-nothing flow from $s$ to $t$ with value $R$ in the constructed network $G$, if and only if the given \textup{NNCCM} has an accepting run.
\end{lemma}

\begin{proof}
Suppose the NNCCM has an accepting run. We assume that we end the run by setting all counters equal to $B$ after the
last check, i.e., by increasing all counters that were smaller than $B$ to $B$.
Fix such an accepting run.
In the remainder of this first part of the proof, the values of counters are as in this accepting run.

First, use in full all edges from $s$ with capacity $L$, and all edges to $t$ with capacity $L+2B$.

It follows from the construction that if at the $t$-th check, counter $j$ has
value $\alpha$, then the outflow of $w_{j,t-1}$ and the inflow of $v_{j,t}$ equal $B+2\alpha$.
Consider such $t$, $j$, and $\alpha$, and suppose that between the $t$-th and $(t+1)$-st check we
increased the counter to $\beta$ (and $\beta = B$ if $t=n$); if the counter was not changed, set $\beta=\alpha$. Thus,
$\beta$ is the value of counter $j$ during the $(t+1)$-st check. 
We now use the following arcs:
\begin{itemize}
    \item $\alpha$ arcs with value 2 from $s$ to $w_{j, 0}$ if $t=1$.  
    \item $\beta-\alpha$ arcs with value 2 from $s$ to $w_{j,t}$.
    \item The arc with value $L+2\alpha$ from $v_{j,t}$ to $w_{j,t}$.
    \item If the $j$th counter is not involved in the $t$th check, or if the firing value of the $j$th counter 
    in the $t$th check unequals $\alpha$, then use the arc from $w_{j,t-1}$ to $v_{j,t}$ with capacity
    $L+2\alpha$.
    \item If the $j$-th counter is involved in the $t$-th check, and the firing value of the $j$-th counter in the $t$-th check equals $\alpha$, then use the arc from $w_{j,t-1}$ to $v_{j,t}$ with capacity
    $L+2\alpha -1$, and the following arcs with capacity $1$: from $w_{j,t-1}$ to $x_{t,1}$, from
    $x_{t,1}$ to $x_{t,2}$, and from $x_{t,2}$ to $v_{j,t}$. Note that, as we have an accepting run,
    the other counter involved in the $t$-th check has the firing value, and thus the edge
    from $x_{t,1}$ to $x_{t,2}$ is used at most once.
\end{itemize}

One verifies that the resulting function is
indeed an all-or-nothing flow. The value equals the inflow at $t$: we have $k$ arcs that send $L+2B$ flow to $t$, to a total value of $R=k\cdot (L+2B)$.

Now, suppose we have an all-or-nothing flow from $s$ to $t$ with value $R$. 

First, look at the arcs out of $s$: we have $k$ arcs with capacity $L$, and $(n+1)kB < L/2$ arcs
with capacity $2$. We must use each of these $k$ arcs with capacity $L$, otherwise the outflow from $s$ and thus
the flow value would be at most $(k-1)L + 2(n+1)kB < kL < R$.

We claim that for each $j\in [1,k]$, the inflow of $w_{j,t}$ is an even number of the form $L+2\alpha$, with
$\alpha \in [0,B]$: we show this by induction w.r.t.\ $t$, the claim being easy for $t=0$. Suppose it holds for $t-1$. 
The node $w_{j,t-1}$ has at most one outgoing edge of capacity $1$, and all other outgoing edges have capacities 
between $L$ and $L+2B$. As $2B<L$, we thus must use exactly one of these `heavy' edges.
Now consider the inflow of $v_{j,t}$. It receives the flow of the used heavy edge and has, in
addition to the heavy edges,
at most one other incoming edge of capacity 1. Hence the inflow
to $v_{j,t}$ is in $[L, L+2B+1]$. As $2B+1<L$, we see that we use exactly one of the outgoing edges
from $v_{j,t}$ --- all have weights in $[L,L+2B]$. The node $w_{j,t}$ has further $B$ incoming edges of weight $2$,
so the inflow of $w_{j,t}$ is in $[L,L+4B]$. It cannot be in $[L+2B+2,L+4B]$, as there is no combination
of edges out of $w_{j,t}$ with total capacity in that interval (using that $4B<L$). Since all arcs
into $w_{j,t}$ have even capacity, the inflow of $w_{j,t}$ is even.

From the above proof, it also follows that the inflow of $w_{j,t+1}$ is at least the inflow
of $w_{j,t}$. Now set the $j$-th counter to $\alpha$ before check $t$ when the inflow of $w_{j,t}$ equals $L+2\alpha$. We see 
that these counters are non-decreasing.

Finally, we verify that the machine does not halt at a check. Suppose that both halves of the $t$-th check
fire, say, for counters $j$ and $j'$, having values $\alpha$ and $\alpha'$. Then 
$w_{j,t-1}$ has an outgoing arc of weight $L+2\alpha-1$, and thus $w_{j,t-1}$ must send  flow
$1$ to $x_{t,1}$. Similarly, $w_{j',t-1}$ has an outgoing arc of weight $L+2\alpha'-1$, and thus $w_{j,t-1}$ must send flow $1$
to $x_{t,1}$. Now $x_{t,1}$ receives 2 inflow, but has only one outgoing arc of capacity 1, which contradicts
the law of flow conservation. Thus, we cannot have a check where both halves fire, and hence  the machine accepts.
\end{proof}

A path decomposition of $G$ of width $O(k)$ can be constructed as follows.
For $0\leq i < n$, we take a bag 

\begin{align*} X_i = \{s,t\} &\cup \{v_{j,i} \mid j\in [1,k]\}
\cup \{v_{j,i+1} \mid j\in [1,k]\}\cup \{w_{j,i} \mid j\in [1,k]\} \\ &\cup \{w_{j,i+1} \mid j\in [1,k]\} \cup \{x_{i+1,1}, x_{i+1,2}\}.
\end{align*}

The size of $G$ is polynomial in $k$ and $n$, and we can construct $G$ with
weight function $c$ and the path decomposition of width $O(k)$ in
$O( k^c \log n)$ space and time, polynomial in $k$ and $n$. Thus, the hardness
result follows.
\end{proof}

\begin{corollary}
The following problems are \textup{XNLP}-complete with pathwidth as parameter:
\begin{enumerate}
    \item {\sc Target Outdegree Orientation}
    \item {\sc Outdegree Restricted Orientation}
    \item {\sc Chosen Maximum Outdegree}
    \item {\sc Minimum Maximum Outdegree}
    \item {\sc Circulating Orientation}
    \item {\sc Undirected Flow with Lower Bounds}
\end{enumerate}
\end{corollary}

\begin{proof}
Membership in XNLP follows easily, with similar arguments as for 
{\sc All-or-Nothing Flow} in the proof of Theorem~\ref{theorem:allornothingxnlp}.

XNLP-hardness for {\sc Target Outdegree Orientation} follows from the hardness
for {\sc All-or-Nothing Flow} (Theorem~\ref{theorem:allornothingxnlp} and the reduction given by Lemma~\ref{lemma:aonf2too}).
XNLP-hardness of {\sc Outdegree Restricted Orientation} follows directly as this problem contains
{\sc Target Outdegree Orientation} as a special case.

XNLP-hardness for {\sc Chosen Maximum Outdegree} follows from the hardness of {\sc Target Outdegree Orientation} by the reduction given by Lemma~\ref{lemma:too2cmo}.
From the hardness of {\sc Chosen Maximum Outdegree}, we obtain by Szeider's transformation (\cite{Szeider11}, see Lemma~\ref{lemma:cmo2mmo}), XNLP-hardness for
{\sc Minimum Maximum Outdegree}.

To prove that {\sc Circulating Orientation} is XNLP-hard, we use the hardness of {\sc Target Outdegree Orientation} and the reduction given by Lemma~\ref{lemma:too2co}. Finally,
as {\sc Circulating Orientation} is a special case of {\sc Undirected Flow with Lower Bounds}, XNLP-hardness of {\sc Undirected Flow with Lower Bounds} follows.
\end{proof}

\subsection{W[1]-hardness for bounded vertex cover number}

In this final section, we consider the complexity of our seven basic orientation and flow problems for a stronger parameter, namely, vertex cover number. We can still prove $\W$-hardness by reduction to the following problem. 

\begin{verse}
{\sc Bin Packing}\\
{\bf Given:}  A set $A=\{a_1, \ldots, a_n\}$ of $n$ positive integers, 
and integers $B$ and $k$ such that $\sum_{i=1}^n a_i = B \cdot k$ \\
{\bf Question:} Partition $A$ into
$k$ sets such that each has sum exactly $B$. 
\end{verse}

Jansen et al.~\cite{JansenKMS13} showed that 
{\sc Bin Packing} parameterized by the number of bins is $W[1]$-hard. 

\begin{theorem}
{\sc Target Outdegree Orientation} parameterized by the size of a vertex cover is $W[1]$-hard.
\label{theorem:targetoutdegreeorientationhard}
\end{theorem}

\begin{proof}
We give a transformation from
{\sc Bin Packing} parameterized by the number of bins.

Suppose $A= \{a_1, \ldots, a_n\}$, $B$ and $k$ are given. We take a complete bipartite graph
$K_{k,n}$. Denote the $k$ vertices at the left colour class by $v_1, \ldots, v_k$
and the $n$ vertices at the right colour class by $w_1, \ldots, w_n$. 

We assign the following weights and target outdegrees. 
All edges incident to $w_i$ have weight $a_i$. Each $w_i$ has as target outdegree $d_{w_i}=a_i$. Each $v_i$ has as target outdegree $d_{v_i}=Bk-B$. 
Let $G$ be the resulting weighted graph.

\begin{lemma}
$A$ can be partitioned into $k$ sets $A_1, \ldots, A_k$ such that each has sum exactly $B$, if and only
if edges in $G$ can be oriented such that the given target outdegrees are fulfilled.
\label{lemma:binpackingorientation}
\end{lemma}

\begin{proof}
Suppose we can partition $A$ into  $k$ sets $A_1, \ldots, A_k$ such that each has sum exactly $B$.
Now construct the folllowing orientation: for each $i\in [1,n]$, $j\in [1,k]$, if $a_i \in A_j$ then orient the edge between $v_j$ and $w_i$ from
$w_i$ to $v_i$; otherwise, orient it from $v_i$ to $w_i$. Each $w_i$ has one outgoing edge of weight
$a_i$. For each vertex $v_i$, the total weight of all incident edges equals $\sum_{i=1}^n a_i = Bk$.
Precisely the edges with weights in $A_i$ are directed to $v_i$ --- as we have a solution of
the {\sc Bin Packing} problem, these have total weight exactly $B$ for each $v_i$. All other edges
are directed out of $v_i$ and have weight $Bk -B$.

Suppose we have an orientation where each vertex has outdegree at most $d_v$. For each $w_i$, exactly
one incident edge must be directed out of $w_i$, as each edge incident to $w_i$ has the same weight, equal to the target outdegree. If the edge between $v_j$ and $w_i$ is directed from $w_i$ to $v_j$, then
assign $a_i$ to $A_j$. Thus, we have that each $a_i$ is assigned to exactly one set $A_j$.
For each $v_j$, the total weight of all edges directed to $v_j$ must be $B$, namely, the total weight
of all incident edges minus the target outdegree. There is a one-to-one correspondence between the weights
of these edges and the numbers assigned to $A_j$, and thus the sum of each $A_j$ equals $B$.
\end{proof}

The result follows from this claim and the quoted result of Jansen et al.~\cite{JansenKMS13}, noting that $K_{k,m}$ has a vertex cover of size $k$.
\end{proof}

\begin{corollary}
The following problems, parameterized by the size of a vertex cover are $W[1]$-hard:
\begin{enumerate}
    \item \textsc{Chosen Maximum Outdegree}
    \item \textsc{Outdegree Restricted Orientation}
    \item {\sc Minimum Maximum Outdegree}
    \item \textsc{Circulating Orientation}
    \item \textsc{Undirected Flow with Lower Bounds}
\end{enumerate}
\end{corollary}

\begin{proof}
Throughout this proof, we look at parameterizations by vertex cover number.

$W[1]$-hardness of \textsc{Chosen Maximum Outdegree} follows from Theorem~\ref{theorem:targetoutdegreeorientationhard}
and the reduction given by Lemma~\ref{lemma:too2cmo}. As \textsc{Outdegree Restricted Orientation} contains
\textsc{Chosen Maximum Outdegree}, it is also $W[1]$-hard.

Lemma~\ref{lemma:cmo2mmo} gives a reduction from \textsc{Chosen Maximum Outdegree} to {\sc Minimum Maximum Outdegree}, 
which gives $W[1]$-hardness of {\sc Minimum Maximum Outdegree}.

The reduction from \textsc{Target Outdegree Orientation} to \textsc{Circulating Orientation} from Lemma~\ref{lemma:too2co}
gives $W[1]$-hardness of \textsc{Circulating Orientation}. \textsc{Undirected Flow with Lower Bounds} contains
\textsc{Circulating Orientation} as a special case, and thus is also $W[1]$-hard.
\end{proof}

We have not yet dealt with {\sc All-Or-Nothing Flow}. It turns out that the NP-completeness proof of {\sc All-Or-Nothing Flow} from Alexandersson~\cite{Alexandersson01}  
also provides a $W[1]$-hardness proof for {\sc All-Or-Nothing Flow} with vertex cover as parameter.
For completeness, we present the argument. 

\begin{theorem}
{\sc All-Or-Nothing Flow} with vertex cover number as parameter is $W[1]$-hard.
\end{theorem}

\begin{proof}
Suppose we have an instance of \textsc{Bin Packing} with $k$ bins of size $B$, and positive integers $a_1, \ldots, a_n$.
Take a graph with vertices $s$, $t$, $v_1, \ldots, v_n$, $w_1, \ldots, w_k$, with arcs $(s,v_i)$ with capacity
$a_i$ ($1\leq i \leq n$), $(v_i,w_j)$ with capacity $a_i$ ($1\leq i\leq n$, $1\leq j\leq k$), and $(w_j,t)$ with capacity $B$ ($1\leq j\leq k$). 
Observe that there is an all-or-nothing flow of value $kB$ in this network if and only if the \textsc{Bin Packing} instance has a solution. Indeed, arguing as in \cite{Alexandersson01}, each vertex $v_i$ receives $a_i$ flow, and must send this
across exactly one outgoing arc to a `bin vertex' $w_j$. Each $w_j$ must receive exactly $B$ flow and send that to $t$.

Note that $\{s, w_1, w_2, \ldots, w_k\}$ is a vertex cover of size $k+1$ of the network.
\end{proof}

\section{Conclusion}

We showed that various classical instances of flow, orientation and capacitated graph problems are \XNLP-hard when parameterized by treewidth (and even pathwidth), but \FPT\ for a novel graph parameter, stable gonality. Following Goethe's motto ``Das Schwierige leicht behandelt zu sehen gibt uns das Anschauen des Unm\"oglichen'', we venture into stating some  open problems. 

\begin{enumerate} 
\item 
Is stable gonality fixed parameter tractable? Can multigraphs of fixed stable gonality be recognized efficiently (this holds for treewidth; for $\sgon=2$ this can be done in quasilinear time \cite{hyperelliptic})? Given the stable gonality of a graph, can a refinement and morphism of that degree to a tree be constructed in reasonable time (the analogous problem for treewidth can be done in linear time)?  Can we find a tree partition of a subdivision with bounded treebreadth? The same question can be asked in the approximate sense. 
\item 
Find a multigraph version of Courcelle's theorem (that provides a logical characterisation of problems that are \FPT\ for treewidth, see \cite{Courcelle}), using stable gonality instead of treewidth: give a logical description of the class of multigraph problems that are \FPT\ for stable gonality.  
\item 
Stable gonality and (stable) treebreadth seems a useful parameter for more edge-weighted or multigraph problems that are hard for treewidth. Find other problems that become \FPT\ for such a parameter. Here, our proof technique of combining tree partitions with ILP with a bounded number of variables becomes relevant. 
\item 
Conversely, find problems that are hard for treewidth and remain hard for stable gonality or (stable) treebreadth. We believe candidates to consider are in the realm of problems concerning ``many'' neighbours of given vertices (where our use of ILP seems to break down), such as \textsc{Defensive Aliance} and \textsc{Secure Set}, proven to be \W-hard for treewidth (but \FPT\ for solution size) 
\cite{BW-SS}, \cite{BW-DA}. For such problems, it is also interesting to upgrade known \W-hardness to \XNLP. 
\item Other flavours of graph gonality (untied to stable gonality) exist, based on the theory of divisors on graphs (cf.\ \cite{Baker}, \cite{BakerNorine}). Investigate whether such `divisorial' gonality is useful parameter for hard graph problems. 
\end{enumerate}

\paragraph*{Acknowledgements} We thank Carla Groenland for various discussions,  and in particular for suggestions related to the capacitated dominating set problems. 
\bibliographystyle{plainurl}

\end{document}